\documentclass[floatfix,aps,twocolumn,floats,pre]{revtex4-1}
\usepackage{graphics,graphicx,epsfig}
\usepackage{epsf,epstopdf,wrapfig}
\usepackage{amsmath,amssymb}
\usepackage{xcolor}
\usepackage{dsfont}

\usepackage{newfloat}
\usepackage{algpseudocode}
\usepackage{dsfont}
\usepackage{algcompatible}
\DeclareFloatingEnvironment[
    fileext=loa,
    listname=List of Algorithms,
    name=ALGORITHM,
    placement=tbhp,
]{algorithm}

\begin{document}
%\vspace*{0.2in}

\title{Adaptive, locally-linear models of complex dynamics}

\author{Antonio C. Costa\textsuperscript{a},
Tosif Ahamed\textsuperscript{b},
Greg J. Stephens\textsuperscript{a,b}}

\affiliation{$^a$Department of Physics and Astronomy, Vrije Universiteit Amsterdam, 1081HV  Amsterdam, The Netherlands\\
$^b$Biological Physics Theory Unit, OIST Graduate University, Okinawa 904-0495, Japan
}

\begin{abstract} 

The dynamics of complex systems generally include high dimensional, nonstationary and nonlinear behavior, all of which pose fundamental challenges to quantitative understanding. To address these difficulties we detail a new approach based on local linear models within windows determined adaptively from data. While the dynamics within each window are simple, consisting of exponential decay, growth and oscillations, the collection of local parameters across all windows provides a principled characterization of the full time series. To explore the resulting model space, we develop a novel likelihood-based hierarchical clustering and we examine the eigenvalues of the linear dynamics.  We demonstrate our analysis with the Lorenz system undergoing stable spiral dynamics and in the standard chaotic regime.  Applied to the posture dynamics of the nematode \emph{C.~elegans} our approach identifies fine-grained behavioral states and model dynamics which fluctuate about an instability boundary, and we detail a bifurcation in a transition from forward to backward crawling.  We analyze whole-brain imaging in {\em C. elegans} and show that global brain dynamics is damped away from the instability boundary by a decrease in oxygen concentration.  We provide additional evidence for such near-critical dynamics from the analysis of electrocorticography in monkey and the imaging of a neural population from mouse visual cortex at single-cell resolution.

\end{abstract}

\maketitle

Complex dynamics are ubiquitous in nature; their diversity in systems ranging from fluids and turbulence \cite{Arratia2005,Majda2014}, to collective motion \cite{Alakent2004} and brain dynamics \cite{Yanagawa2013}, is unified by common challenges of analysis which include high dimensionality, nonlinearity and nonstationarity. But how do we capture the quantitative details of the dynamics of complex systems with models simple enough to offer substantial intepretability?

Motivated by the remarkable increase in data quantity and quality as well as growing computational power, one approach is to fit a single global model to the dynamics with properties extracted from data. For example, deep neural networks and other machine learning techniques \cite{Li2017,Pathak2017} often produce high dimensional nonlinear models, which can precisely represent complex dynamics and yield accurate predictions. While powerful however, these methods can create representations of the dynamics that are too intricate for simple conceptual understanding. Another approach uses sparse regression to find a system of differential equations governing a nonlinear dynamical system \cite{Brunton2016a}. Also, short time brain oscillations were studied using jPCA \cite{Churchland2012}, a method that approximates the dynamics as a linear model with skew-symmetric couplings. Although promising, global methods are unable to handle non-stationarities, such as when a time series is composed of a set of distinct dynamics that change in time.

An alternative to global methods is to segment the dynamics into simpler components which change in time. For example, a low-dimensional representation of the spatiotemporal patterns found in the human brain was obtained through dynamic mode decomposition \cite{SCHMID2010} in short temporal segments \cite{Brunton2016}. Studies on self-regulated dynamical criticality in the human brain used vector autoregressive models locally in time \cite{Solovey2012}. Behavioral motifs in \emph{Drosophila melanogaster} were found using local-time wavelet analysis \cite{Berman2014a}. In these methods however, the local windows are defined phenomenologically, which may conflate distinct dynamical behaviors.

Principled approaches for the segmentation of time series include those of change-point detection \cite{Guralnik1999,Takeuchi2006,Wang2012,Liu2013,Chen2013,Omranian2015,Preus2014,Kawahara2007}, which aim to identify structural changes in the time series but often focus on the location of change-points or forecasting, instead of the underlying dynamics \cite{Chen2013,Wang2012,Preus2014,Liu2013,Kawahara2007,Omranian2015}.  Other techniques such as hidden Markov models \cite{Bryan2015,Stanculescu2014,Gallagher2013}, assume that the global dynamics are composed of a set of underlying dynamical states which the system revisits, without providing a parameterization of the underlying dynamical patterns \cite{Gallagher2013}. More recently, switching linear dynamical systems and autoregressive hidden Markov models  \cite{Wiltschko2015,Linderman2016,Markowitz2018} were developed with the aim of providing such a parameterization, but they do so either by setting the number of breaks from the onset \cite{Guthery1974,Society1976}, or by assuming that there is a set of underlying dynamical regimes and that the system switches between them \cite{Chamroukhi2013,Linderman2016,Stanculescu2014,Bryan2015,Wiltschko2015,Markowitz2018}.

Here, we combine the simplicity of linear dynamical systems with a likelihood-based algorithm for identifying dynamical breaks to construct interpretable, data-driven models of complex dynamics, with minimal \emph{a priori} assumptions about the breakpoints or the number of states.  We approximate the full dynamics with first-order linear dynamical systems in short windows and use a likelihood-ratio test to estimate to what extent newly added observations fit the same linear model, thus adaptively determining the size of the local windows. The global dynamics is therefore parameterized as a set of linear couplings within windows of various lengths. We analyze the resulting model space using hierarchical clustering with a new likelihood-based similarity measure and by examining the dynamical eigenvalue spectra in three illustrative systems: the Lorenz dynamical system and both posture and whole brain dynamics of the nematode \emph{C.~elegans}.  In addition, we extend our analysis to higher-dimensional dynamics: electrocorticography (ECoG) recordings in non-human primates and a population of hundreds of neurons in mouse visual cortex.

\section*{Locally-linear, adaptive segmentation technique}

\begin{figure*}
\centering
\includegraphics{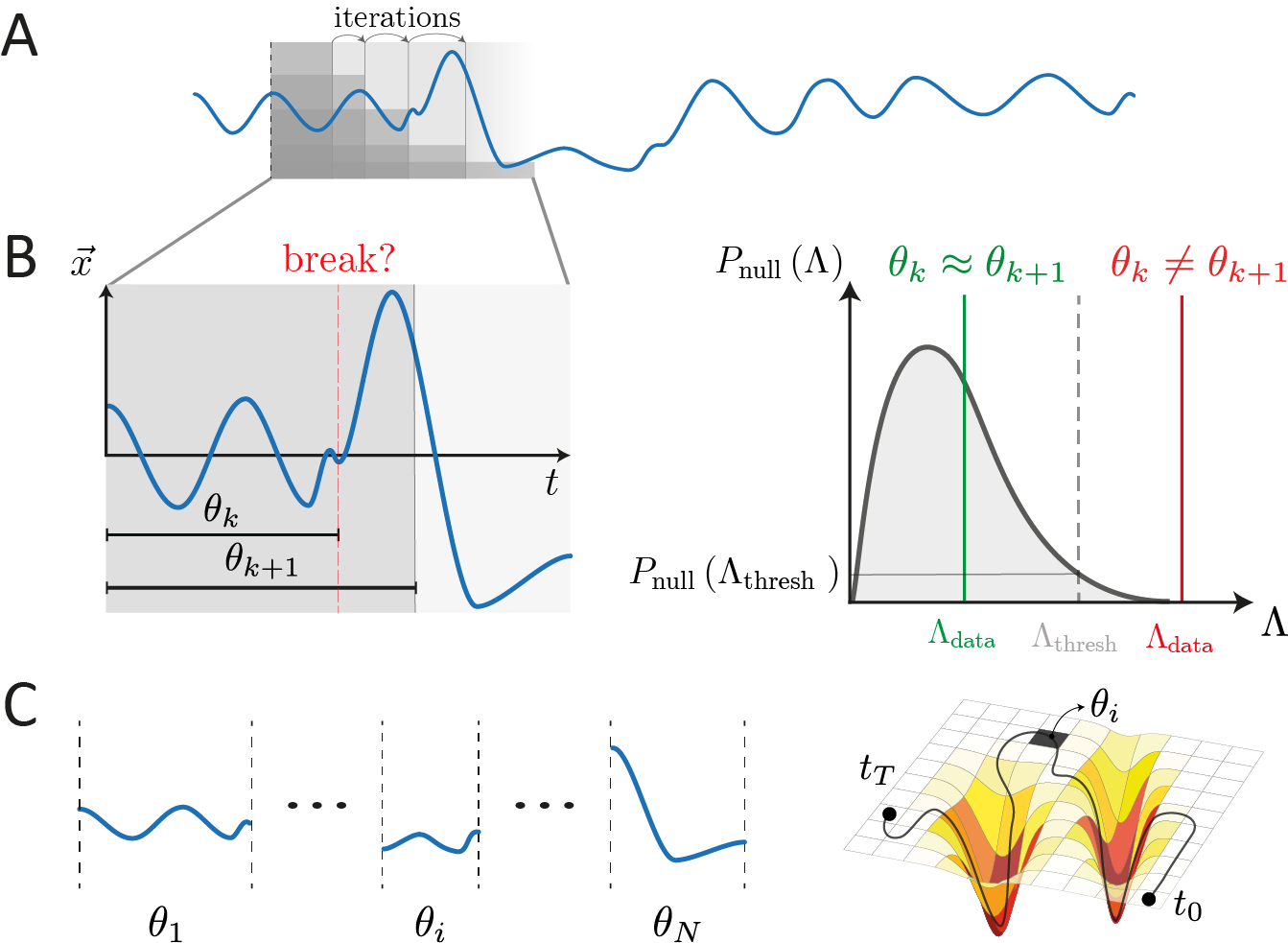}
\caption{{\bf Schematic of the adaptive, locally-linear segmentation algorithm.} 
(A) A $d$-dimensional time series is depicted as a blue line. We iterate over pairs of subsequent windows and use a likelihood-ratio test to assess whether there is a dynamical break between windows. 
(B) We compare linear models $\theta_{k}$ and $\theta_{k+1}$, found in the windows $X_k$ and $X_{k+1}$, by the log-likelihood ratio $\Lambda_\text{data}$, \eqref{SEq:LogLikRatio}. To assess significance we compute the distribution of log-likelihood ratios under the null hypothesis of no model change $P_{\rm null}(\Lambda)$ and identify a dynamical break when $\Lambda_{\rm data} > \Lambda_\textrm{thresh}$ where $P_{\rm null}(\Lambda_\textrm{thresh})=0.05$. If no break is identified, we continue with the windows $\{\theta_{k+1},\theta_{k+2}\}$. (C) The result of the segmentation algorithm is a set of windows of varying lengths and model parameters $\left\{\theta_1,\ldots,\theta_N\right\}$. Our approach is similar to approximating a complex-shaped manifold by a set of locally flat patches, and encodes a nonlinear time series through a trajectory within the space of local linear models.
}
\label{fig:Panel1}
\end{figure*}

An overview of the segmentation technique is given in Fig.\,(\ref{fig:Panel1}) and a detailed description as well as links to publicly available code are in Methods. Briefly, we iterate over pairs of consecutive windows, Fig.\,(\ref{fig:Panel1}A), and estimate whether the linear model fit in the larger window $\theta_{k+1}$ is significantly more likely to model the observations in the larger window when compared to the model found in the smaller window $\theta_k$, Fig.\,(\ref{fig:Panel1}B). We compare the two models by the log-likelihood ratio $\Lambda_\text{data}$ and assess the significance of $\Lambda_{\rm data}$ by using Monte Carlo methods to construct a likelihood ratio distribution $P_{\rm null}(\Lambda)$ under the null hypothesis of no model change. This null distribution is used to define $\Lambda_{\rm thresh}$ according to a threshold probability, or significance level $P_{\rm null}(\Lambda_\textrm{thresh})$. We identify a dynamical break when $\Lambda_{\rm data} > \Lambda_{\rm thresh}$ in which case we save the model parameters and start a new modeling process from the break location. If  $\Lambda_{\rm data} \leq \Lambda_{\rm thresh}$ then no break is identified and we move to the next window pair $ \{\theta_{k+1}, \theta_{k+2}\}$.  Over the entire time series,  our procedure yields a set of $N$ windows of varying sizes with their respective linear model parameters, $\left\{\theta_1,\ldots,\theta_N\right\}$, Fig.\,(\ref{fig:Panel1}C) and is analogous to tiling a complex shaped manifold into local flat regions. We thus trade the complexity of the nonlinear time series for a space of simpler local linear models that captures important properties of the full dynamics.

\section*{Surveying the space of models}
The application of locally-linear, adaptive segmentation generally results in a large set of linear dynamical systems (LDS) and we explore this space both through the eigenvalues of the coupling matrices and by model clustering through a novel, likelihood-based measure of similarity. Despite a typically large number of models, the dynamical eigenvalues offer a direct measure of local oscillations and stability. Complex conjugate eigenvalues represent oscillations, with frequency $f=\rm{Im}(\lambda)/(2\pi)$. A negative real part implies stable damped dynamics along that mode, while a positive real part implies unstable exponentially growing trajectories. As the least stable eigenvalue approaches 0 the system becomes sensitive to external perturbations. At the bifurcation point, $\rm{Re}(\lambda)=0$, the susceptibility diverges and we enter a critical dynamical state \cite{Munoz2018,Magnasco2009a,Magnasco2009}. The full spectrum of eigenvalues across models thus provides not only information about oscillatory patterns but also stability and criticality.

To cluster the models, we note that simply using the Euclidean metric is inappropriate, since the space of linear models is invariant under the action of the GL($n$) group \footnote{The action of $P \in \textrm{GL}(n)$ to the matrix of linear couplings $A$ results in new coupling matrix $PA$ that is very different according to the euclidean metric, while representing the same linear dynamics. Therefore, using the Euclidean distance is deeply misleading as two matrices that are distant in euclidean metric can represent the same linear dynamical system.}. Instead, we define dissimilarity as the loss in likelihood when two windows are modeled by a single linear model constructed fitting within the combination of windows. Given two windows, $X_a$ and $X_b$, we define the dissimilarity as $d_{a,b} = \Lambda_{c,a} + \Lambda_{c,b}$, where $\Lambda$ is the log-likelihood ratio and $c$ is the union of the windows $X_c=X_a\cup X_b$.  We note that this measure is symmetric $d_{a,b}=d_{b,a}$, positive semi-definite $d_{a,b}\geq 0$ and does not require the windows to be the same size. If the dynamics in both windows are similar, then the combined model will still accurately fit each window. If not, then it will be far less likely to model the windows, resulting in a higher disparity between models. Once the dissimilarity is computed between all models we perform hierarchical clustering by combining models according to Ward's minimum variance criterion \cite{Ward1963}.

\section*{Lorenz system}
\begin{figure*}
\begin{center}
\includegraphics{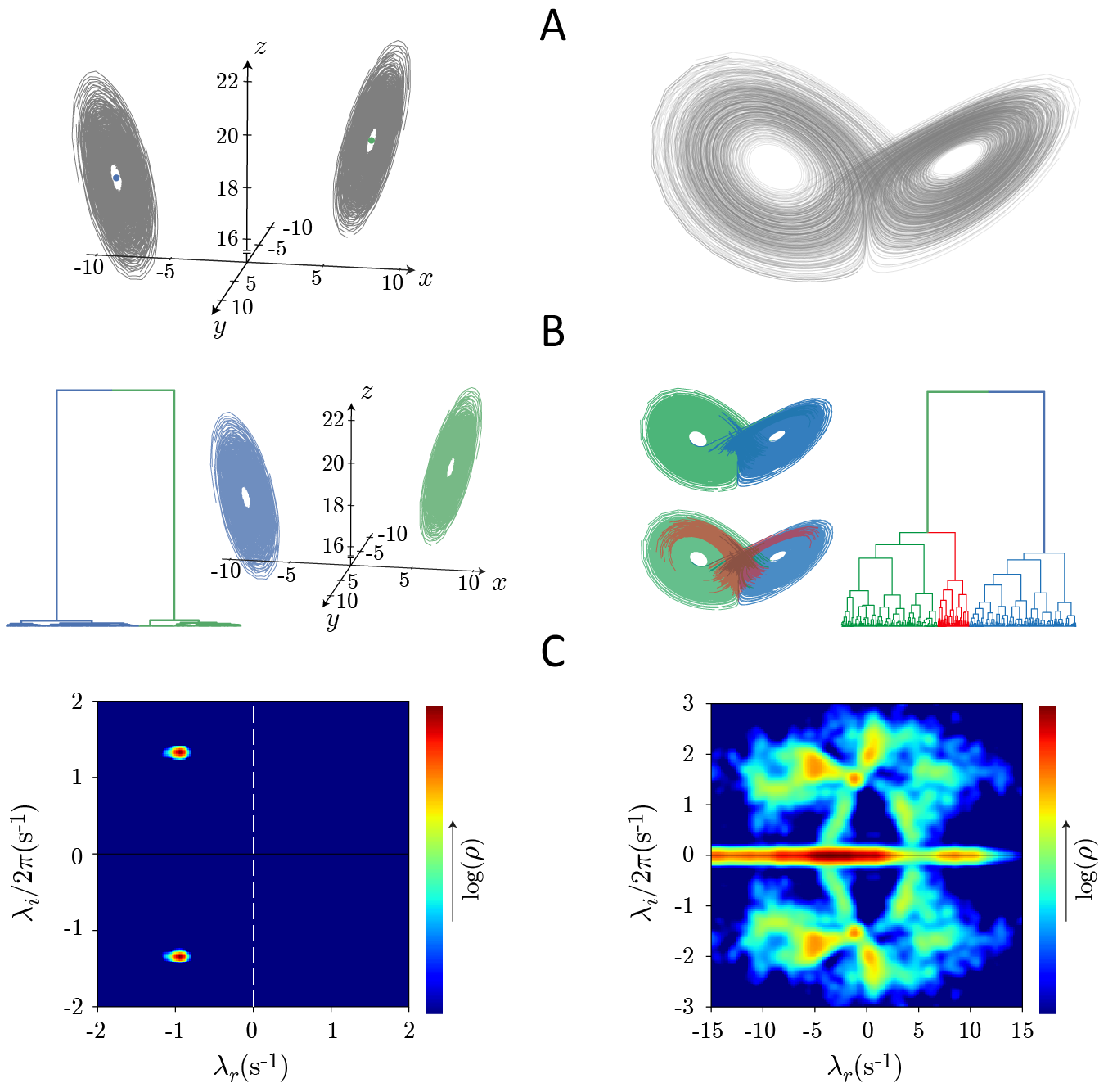}
\caption{{\bf Adaptive segmentation of the Lorenz dynamical system and likelihood-based clustering of the resulting model space.}  (A) Simulated Lorenz system for stable spiral dynamics (left) $\{\rho=20, \beta=8/3, \sigma=10\}$ and the standard chaotic regime (right) $\{\rho = 28, \beta=8/3, \sigma = 10\}$. (B) Likelihood-based hierarchical model clustering. In the spiral dynamics there is a large separation between models from each lobe, while the dynamics within lobe are very similar. In the chaotic regime, the model-space clustering first divides the two lobes of the attractor and the full space is intricate and heterogeneous.  (C) Dynamical eigenvalue spectrum for each regime, $\lambda_r$ and $\lambda_i$ respectively represent the real and imaginary eigenvalues. The spiral dynamics (left) exhibits a pair of stable, complex conjugate peaks while in the chaotic regime (right) we find a broad distribution of eigenvalues, often unstable, reflecting the complexity of the chaotic attractor. 
}
\label{fig:Panel2}
\end{center}
\end{figure*}

As a demonstration of the locally-linear approach, we analyze the time series generated from the Lorenz dynamical system \cite{Lorenz1963}:

\begin{align}
\dot{x} = &\,\sigma(y-x) \nonumber \\ 
\dot{y} = &\,x(\rho-z)-y \nonumber \\
\dot{z} = &\,xy-\beta  z,\,  \nonumber 
\end{align}
with $\beta = 8/3$ and $\sigma = 10$.  We explore two dynamical regimes: transient chaos with late-time, stable spiral dynamics at $\rho=20$ and the standard chaotic attractor with $\rho=28$. For spiral dynamics, we vary the initial conditions to sample the dynamics approaching the fixed point at the center of each lobe, Fig.\,(\ref{fig:Panel2}A), which have the same period but vary in their phase space trajectories. We apply adaptive segmentation and show the result of model-space clustering in Fig.\,(\ref{fig:Panel2}B). We find a single dominant split in the clustering dendrogram, which corresponds to approaching the two different fixed points. Inside each branch the different linear models are all quite similar. In the chaotic regime however, we find substantially more structure and large dissimilarities between models even at the lower branches of the tree. Notably, the first split occurs between the two lobes of the attractor and more generally, the linear model clustering provides a partition of the Lorenz phase space with different levels of description depending on the depth in the dendrogram.

Further insight into the dynamics is reflected in the distribution of the spectrum of eigenvalues across the local linear models, Fig.\,(\ref{fig:Panel2}C). In the spiral dynamics, we find two peaks reflecting a dominant pair of complex conjugate eigenvalues and these correspond to a decaying oscillation ($\operatorname{Re}{(\lambda)}<0$). We note that while the local coupling matrix is constructed from finite temporal windows and is not the instantaneous Jacobian, the dynamical eigenvalues are close to those derived from linear stability of the fixed points (Methods).  In contrast, the spectrum in the chaotic regime reflects a complexity of behaviors, with many models displaying unstable dynamics along the $1d$ unstable manifold of the origin, Fig.\,(S1). In the locally-linear perspective, the complexity of chaotic dynamics is associated with both substantial structure in the space of models as revealed through hierarchical clustering, as well as a wide range of dynamics, including eigenvalues that are broadly distributed across the instability boundary.

\section*{Posture dynamics of \emph{C.~\MakeLowercase{elegans}}}

\begin{figure*}
\begin{center}
\includegraphics{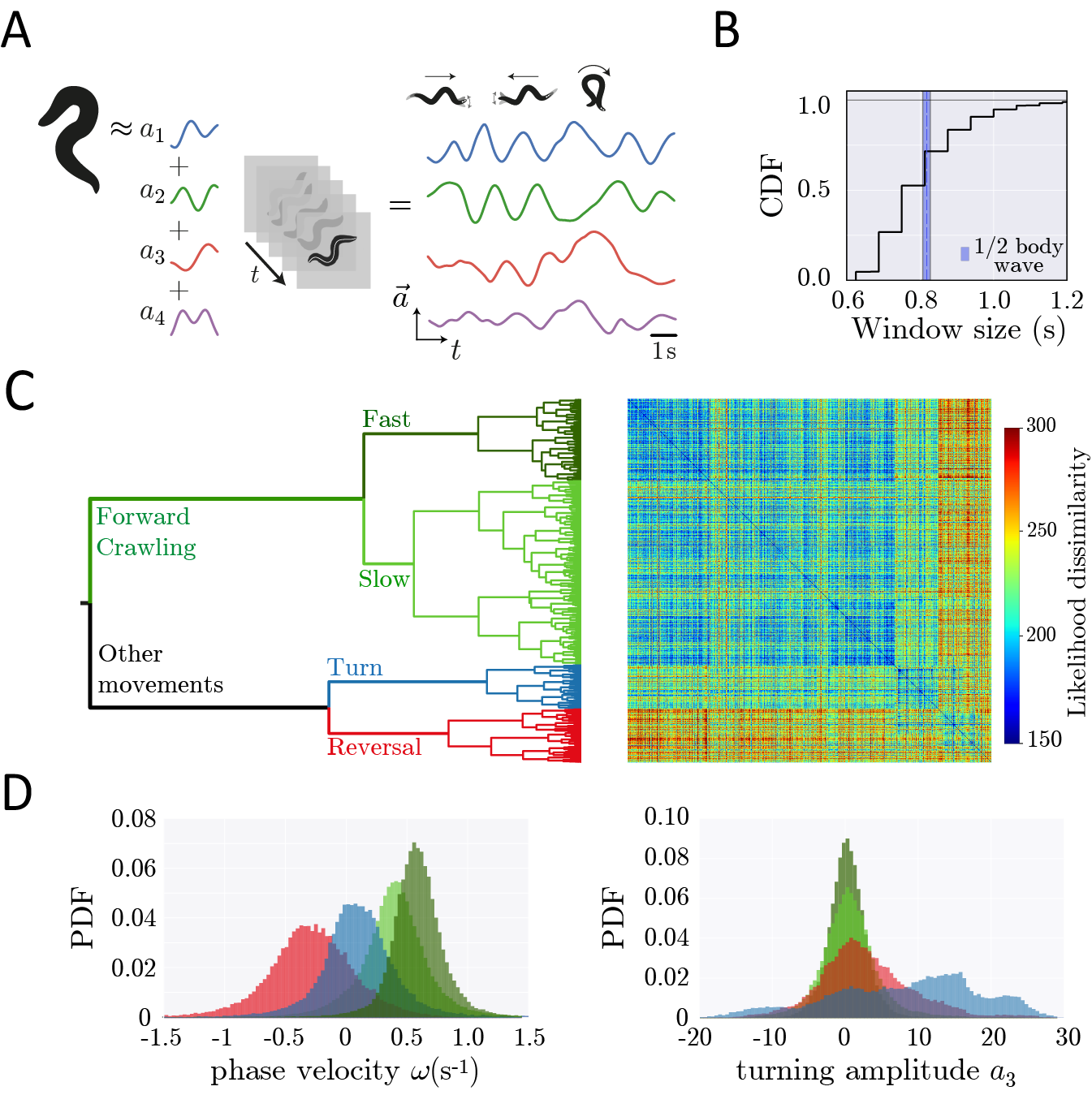}
\caption{{\bf Locally-linear analysis of \emph{C.~elegans} posture dynamics reveals a rich space of behavioral motifs.} (A) We transform image sequences into a 4D posture dynamics using  ``eigenworm''  projections \cite{Stephens2008} where the first two modes $(a_1,a_2)$ describe a body wave, with positive phase velocity $\omega$ for forward motion and negative $\omega$ when the worm reverses. High values of $|a_3|$ occur during deep turns, while $a_4$ captures head and tail movements. (B) The cumulative distribution (CDF) of window sizes reveals rapid posture changes on the timescale of the locomotor wave (the average duration of a half body wave is shown for reference). (C) Likelihood-based hierarchical clustering of the space of linear posture dynamics. At the top of the tree, forward crawling models separate from other behaviors. At the next level, forward crawling splits into fast and slower body waves, while the other behaviors separate into turns and reversals. Hierarchical clustering results in a similarity matrix with weak block structure; while behavior can be organized into broad classes, large variability remains within clusters. (D) Cluster branches reveal interpretable worm behaviors. We show the probability distribution (PDF) of body wave phase velocities and turning amplitudes at the 4-branch level of the tree. In the first forward state (dark green) worms move faster than in the second branch (light green). In the turn branch (blue), the phase velocity is centered around zero and high values of $|a_3|$ indicate larger turning amplitudes. In the reversal branch (red) we find predominantly negative phase velocities.
}

\label{fig:Panel3}
\end{center}
\end{figure*}

The posture dynamics of the nematode \emph{C.~elegans} is accurately represented by a low dimensional time series of ``eigenworm'' projections \cite{Stephens2008}, Fig.\,(\ref{fig:Panel3}A), though a quantitative understanding of the behaviors in these dynamics remains a topic of active research \cite{Brown2013,Yemini2013,Broekmans2016,Liu2018}. More broadly, principled behavioral analysis is the focus of multiple recent advances in the video imaging of unconstrained movement across a variety of organisms \cite{Berman2014a,Berman2016,Berman2018,Klibaite2018,Calhoun2017,Wiltschko2015,Han2018,Szigeti2015,Todd2017,Bruno2017}. Here, we apply adaptive locally-linear analysis to the eigenworm time series and find short model window lengths ranging from approximately $0.6\,{\rm s}$ to $1.2\,{\rm s}$, Fig.\,(\ref{fig:Panel3}B). Notably the median model window size is similar to the duration of half a worm's body wave suggesting that the body wave dynamics provide an important timescale of movement control.

Likelihood-based model clustering reveals that forward crawling separates from other worm behaviors at the top level of the hierarchy, Fig.\,(\ref{fig:Panel3}C). At a finer scale, forward crawling breaks into faster and slower models, while turns and reversals emerge from the other branch.  To clarify the structure of the model space we leverage the interpretability of the eigenworm projections where the first two modes ($a_1$ and $a_2$) capture a primary body wave oscillation with phase velocity $\omega=-\frac{d} {dt} \tan^{-1}{(a_2/a_1)}$  while a third projection $a_3$ captures broad body turns \cite{Stephens2008}. In Fig.\,(\ref{fig:Panel3}D) we show $\omega$ and $a_3$ for each cluster. We note that there are a few low amplitude positive phase velocities in the reversal branch: the adaptive segmentation detects a dynamical break when the worm starts slowing down in preparation for a reversal, and those first frames are included in the reversal window. We note that changes in the activity of AIB, RIB and AVB neurons also precede the reversal event \cite{Kato2015}. Further examination of the agreement between model clusters and behavioral states is provided in in Fig.\,(S2). At a coarse level, the canonical behavioral states described since the earliest observations of the movements of  \emph{C.~elegans} \cite{Croll1975,Schwarz2015} are identified here using data-driven, quantitative methods. 

The model parameters provide an additional opportunity for interpretation of the worm's behavior and in Fig.\,(S3) we show the coefficients for illustrative models at the clustering level consisting of four states. For models from the two forward states, the two pairs of complex conjugate eigenvalues have different imaginary values, corresponding to different frequencies of the locomotor wave oscillation. On the other hand, the turning model can be identified by the large mean turning amplitude. Finally, the reversal model exhibits an inversion in the sign of the $\{a_1, a_2\}$ coupling, which corresponds to a reversal in the direction of the body wave.

The full structure of the model dendrogram reveals that the behavioral repertoire of {\it C.~elegans} is far more complicated than the canonical states of forward, reversal and turning locomotion. For example, forward crawling behavior is rich and variable: two forward crawling models can be almost as dissimilar as a turn is from a reversal. While the worm's behavior is stereotyped at a coarse-grained level, there is significant variation within each of the broad behavioral classes. For example at the 12-branch level of the tree, the reversal class splits into faster and slower reversals as well as new behavioral motif: a reversal-turn, Fig.\,(S4).  Certainly, some of these ``states'' simply reflect the linear basis of the segmentation algorithm.  However, longer nonlinear behavioral sequences can emerge from analysis of the resulting symbolic dynamics.

\begin{figure*}
\begin{center}
\includegraphics{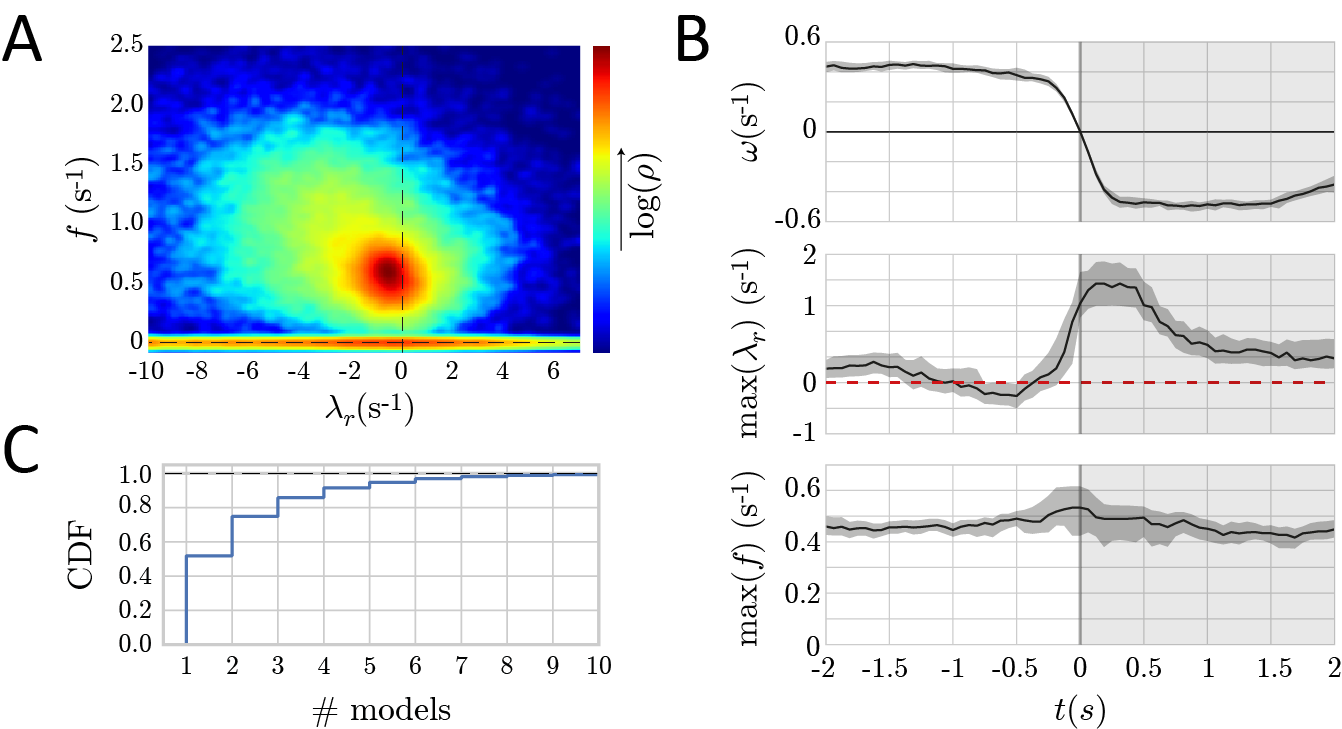}
\caption{{\bf Linear posture dynamics in {\em C.~elegans} is distributed across an instability boundary with spontaneous reversals evident as a bifurcation.} (A) The eigenvalues of the segmented posture time series reveal a broad distribution of frequencies $f=|\text{Im}(\lambda)|/2\pi$ with a peak $f\sim 0.6\, {\rm s}^{-1}$ that spills into the unstable regime. (B) We align reversal events and plot the maximum real eigenvalue ($\lambda_r$) and the corresponding oscillation frequency. As the reversal begins, the dynamics become unstable, indicating a Hopf-like bifurcation in which a pair of complex conjugate eigenvalues crosses the instability boundary. The shaded region corresponds to a  bootstrapped 95\% confidence interval. (C) Instabilities are both prevalent and short-lived. We show the cumulative distribution of the number of consecutive stable or unstable models demonstrating that bifurcations also occur on short-times between fine-scale behaviors.
}
\label{fig:Panel4}
\end{center}
\end{figure*}

We analyze the spectrum of eigenvalues across the entire model space, Fig.\,(\ref{fig:Panel4}A), and find that the worm's dynamics includes both stable and unstable eigenvalues with a broad peak at $f \sim 0.6\,{\rm s}^{-1}$, in agreement with the average forward undulatory frequency of the worm in these food-free conditions \cite{Stephens2008}. Some of these unstable dynamics are explained by coarse behavioral transitions and we align reversal trajectories by the moment when the body wave phase velocity $\omega$ crosses zero from above in order to follow the median of the least stable eigenvalues during this transition, Fig.\,(\ref{fig:Panel4}B).  We see that the reversal behavior is accompanied by an apparent Hopf bifurcation: a pair of complex conjugate eigenvalues crosses the instability boundary.  More generally, we find that the dynamics rapidly switches stability, Fig.\,(\ref{fig:Panel4}C).  Indeed the spectrum of eigenvalues shows that the worm's dynamics is generically near the instability boundary which is suggestive of a general feature of flexible movement control. 

\section*{Neural dynamics of \emph{C.~\MakeLowercase{elegans}}}
With recent progress in neural imaging, {\em C.~elegans} also provides the opportunity to observe whole-brain dynamics at cellular resolution \cite{Nguyen2015,Kato2015,Nichols2017,Schrodel2013,Prevedel2014,Venkatachalam2015} and we apply our techniques to analyze the differences between active and quiescent brain states driven by changes in oxygen ($\text{O}_2$) concentration \cite{Nichols2017}. In these experiments,  worms enter a ``sleep''-like state when the $\text{O}_2$ levels are lowered to $10\%$, and were aroused when the $\text{O}_2$ concentration is increased to $21\%$.  These conditions offer a probe of the neural dynamics of {\em C. elegans} and also suggest qualitative comparisons with sleep transitions measured through electrocorticography (ECoG) in human and non-human primates \cite{Solovey2012,Solovey2015,Alonso2014}.

\begin{figure*}
\begin{center}
\includegraphics{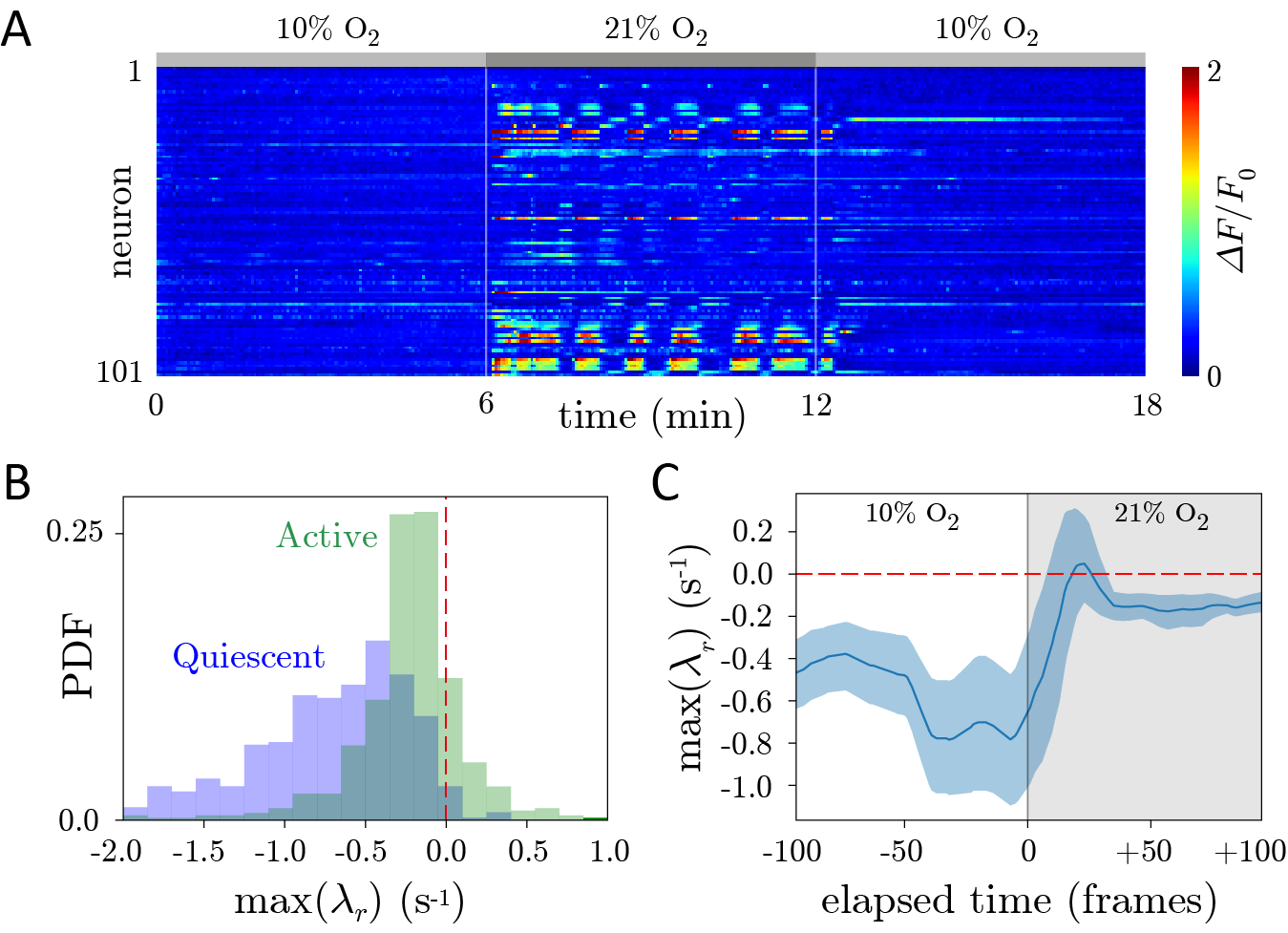}
\caption{{\bf Quiescence stabilizes global brain dynamics in {\em C. elegans}}.
(A) We analyze whole-brain dynamics from previous experiments in which worms were exposed to varying levels of $\text{O}_2$ concentration \cite{Nichols2017}.  We show the background subtracted fluorescence signal $\Delta F/F_0$ from 101 neurons while $\text{O}_2$ concentration changed in 6 minute periods: low $\text{O}_2$ (10\%) induces a quiescent state; high $\text{O}_2$ (21\%) induces an active state. 
(B) We plot the distribution of maximum real eigenvalues ($\lambda_r$) for the active and quiescent states. The active state is associated with substantial unstable dynamics,  while the dynamics of the quiescent state is predominately stable which is consistent with putative stable fixed point dynamics. (C) We plot the average maximum real eigenvalue as the $\text{O}_2$ concentration is changed. We align the time series from different worms to the first frame of increased $O_2$ concentration and show the accompanying increase in the maximum real eigenvalue, which crosses and remains near to the instability boundary. The shaded region corresponds to a bootstrapped 95\% confidence interval and curves were smoothed using a 5-frame running average.
}
\label{fig:Panel5}
\end{center}
\end{figure*}

In Fig.\,(\ref{fig:Panel5}A) we show an example trace of the recorded neural activity and further details are available in Methods. 
We analyze the stability of the neural dynamics using ``active'' and ``quiescent'' global brain states identified previously \cite{Nichols2017} and we show the distribution of least-stable dynamical eigenvalues for each condition, Fig.\,(\ref{fig:Panel5}B). To further characterize the transition between states, we align the maximum real dynamical eigenvalues by the time of increased $\text{O}_2$ concentration and show the mean of this distribution, Fig.\,(\ref{fig:Panel5}C). As activity increases from the quiescent state, the dynamics move towards the instability boundary, eventually crossing and remaining nearly unstable in the aroused state. While the neural imaging occurred in paralyzed worms, the broad distribution of eigenvalues across the instability boundary in the active brain state is consistent with the complexity of the behavioral dynamics. Notably, the model space also contains clusters in approximate correspondence with previous state labels, both in these experiments \cite{Nichols2017} and in worms exhibiting more complex natural behaviors \cite{Kato2015}, Fig.\,(S5) and Fig.\,(S10).

\section*{Higher dimensional systems}

\begin{figure*}
\begin{center}
\includegraphics{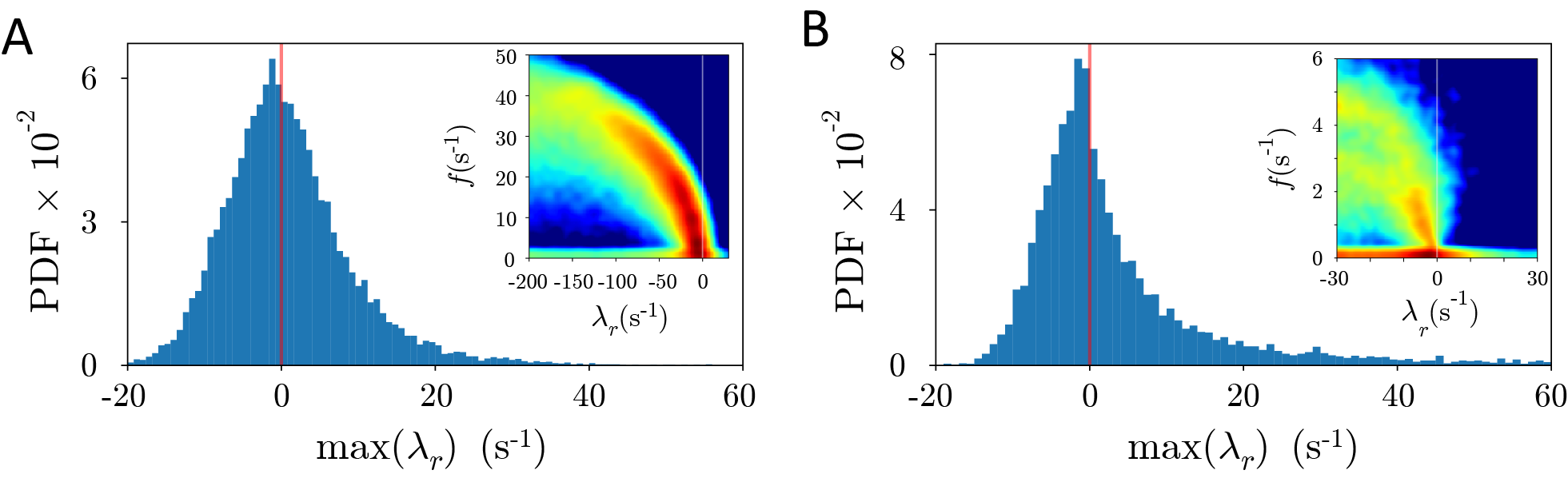}
\caption{{\bf Higher dimensional applications of the adaptive locally-linear model technique: the dynamics exhibit a wide range of frequencies and near-critical behavior.} 
(A) Distribution of the least stable real eigenvalues from each window of the local linear models obtained from the analysis of ECoG recordings in non-human primates. The inset shows the full distribution of eigenvalues - color code is the same as in Fig.\,(\ref{fig:Panel3}). 
(B) Distribution of the least stable real eigenvalues from each window of the local linear models obtained in recordings of 240 neurons in the visual cortex of \emph{Mus musculus}. The inset shows the full distribution of eigenvalues - color code is the same as in Fig.\,(\ref{fig:Panel3}). Here, due to the high-dimensionality, a regularization procedure was added to the original technique (Methods).
}
\label{fig:Panel6}
\end{center}
\end{figure*}

Beyond the previous examples, there are situations where high dimensionality and low sampling rate (relative to the signal correlation time) yield a minimum window size which is too large to capture important dynamics. This is as expected--more dimensions generally require more statistical samples--and our minimum window size is chosen conservatively to result in a good model fit without regularization and thus without bias. 
If the sampling rate is adequate then we can easily apply our technique to higher dimensional data, as we demonstrate in  Figs.\,(\ref{fig:Panel6}A,\,S6) where we show the analysis of 40 components from ECoG recordings in non-human primates. For sparsely sampled systems on the other hand, regularization is generally required to accurately compute the inverse of the data and error covariance matrices.  We offer one straightforward procedure which is motivated by Principal Component Regression \cite{Jolliffe1982} where we reduce the dimension locally, within each window. We detail this idea in Methods and provide a demonstration from recordings of hundreds of neurons in mouse visual cortex Figs.\,(\ref{fig:Panel6}B,\,S7).  In both of of these high-dimensional systems, the local-linear analysis  yields model dynamics that sit near the instability boundary, with a large fraction of unstable models, Fig.\,\ref{fig:Panel6}.

\section*{Discussion}

Simple linear models form the foundation for our analysis of complex time series based upon interpretable dynamics in short segments determined adaptively from data. The trajectories of a single model can only exponentially grow, decay or oscillate.  Yet, by tiling the global dynamics with many such models we faithfully reproduce nonlinear, multidimensional and nonstationary behavior and parameterize the full dynamics with the set of local couplings. To elucidate the resulting space of models, we constructed hierarchical clusters with a new likelihood-based dissimilarity measure between local dynamics, and we examined the distribution and stability of the dynamical eigenvalues.

In the Lorenz system, chaos is distinguished by an increased model variety, including many with instabilities. In the chaotic attractor, the model hierarchy naturally splits across the two lobes with the clusters at deeper levels forming a progressively finer partition of the phase space.  These partitions, as well as the recurrence structure in the space of models, can be used to estimate ergodic properties of the attractor such as the Kolmogorov-Sinai entropy \cite{Kolmogorov1959,ott2002chaos}. 
Adaptive locally-linear analysis offers a new approach for thinking quantitatively about animal behavior, where recent advances have resulted in multiple efforts aimed at understanding movement at high resolution \cite{Stephens2008, Brown2013, Berman2014a}. In the posture dynamics of {\em C.~elegans} we found that interpretable behavioral motifs emerged naturally, with high-level clusters reflecting canonical behavioral states of forward, reversal and turning locomotion \cite{Croll1975} and finer-scale, novel states appearing deeper in the tree. An advantage of our clustering approach is that the level of behavioral description can be chosen appropriate to the nature of the analysis and these states form a natural basis with which to apply techniques such as compression \cite{Brown2013,Gomez-Marin2016} and to explore long-time behavioral dynamics like memory \cite{Berman2014a}. The dissimilarity measure also enables the comparison of models across datasets, regardless of experimental details such as frame rate, as long as postures are projected into the same basis. This can be useful for developing a master repertoire of behaviors \cite{Brown2013} as well as looking for differences between nematode species or studying perturbations to behavior \cite{Gomez-Marin2016,Schwarz2015,Stephens2008,Broekmans2016,Vidal-Gadea2011,Gao2018,Fouad2018}. We note that the success of the local linear basis in revealing interpretable worm behavior results in part from the ability to capture oscillations and the interactions between different posture modes, both common components of movement behavior. 

The eigenvalues of the posture dynamics reflect variability and hint at the presence of flexible control. While the eigenvalue distribution is centered on the frequency of the locomotor wave, the peak is close to the instability boundary and many models are unstable. Posture movements thus appear more complex than suggested by a model of stereotyped behaviors composed of a small collection of simple limit cycles \cite{Revzen2012}.

The global neural activity of {\em C. elegans} also displays model dynamics which fluctuate across the instability boundary Fig.~(\ref{fig:Panel5}B), suggesting a near-critical brain state (see \cite{Chen2019} for a similar, recent conclusion from a statistical perspective). Additionally, we obtained similar findings through local linear analysis of ECoG in monkey and single-cell recordings from a neural population in mouse visual cortex, Fig.\,(\ref{fig:Panel6}). Such behavior was previously observed in whole brain activity \cite{Solovey2012,Solovey2015} and is consistent with the observation that the firing rate of neural populations exhibits subcritical dynamics \cite{Wilting2018}. Dynamical criticality is advantageous for information processing in models of neural networks \cite{Toyoizumi2011,Sussillo2009} and can occur as a result of an anti-Hebbian balance of excitation and inhibition \cite{Magnasco2009}. Close to criticality, the dynamics is highly susceptible to external perturbations and small changes to the stability can have a dramatic impact on the dynamical time scales \cite{Wilting2018b}. This susceptibility can change across brain regions \cite{Murray2014} and we show that it can also be modulated with behavioral transitions and neural quiescence in \emph{C. elegans}. Such modulation also occurs with the induction of anesthesia in ECoG \cite{Alonso2014}, Fig.\,(S6).

For simplicity and interpretability, we chose a basis of first-order linear models, though extensions to higher order are straightforward. Also, while we have focused on the deterministic model properties, the error terms (\eqref{SEq:VAR}, Methods) may also carry important information. For example, it has been recently shown that even deterministic chaotic systems can be accurately represented as linear dynamics with a heavy-tailed stochastic forcing, the magnitude of which can be used to identify bursting or lobe switching events \cite{Brunton2017}. In our analysis we find that the error distribution exhibits heavy tails along the direction of the nonlinearities of the Lorenz system, and that the magnitude increases with lobe switching in the Lorenz system or reversal events in \emph{C.~elegans}.

There have been multiple recent advances in applying linear models to the analysis of complex time series \cite{Oh2008,Fox2009,Linderman2016,Wiltschko2015,Markowitz2018} and while our approach shares a linear basis, there are important differences.  For example, both autoregressive hidden Markov models and switching linear dynamical systems assume that the dynamics is composed of a set of discrete coarse-grained dynamical modes, revisited by the system. The number of these modes is a hyperparameter of the model, chosen to balance model complexity and accuracy.  In contrast, our analysis finds as many linear models as permitted by reliable estimation and the depth of the hierarchical clustering can be chosen {\em a posteriori} depending on the interpretation of the clusters. Our combination of adaptive segmentation and hierarchical clustering also enables the explicit examination of the variability of models within each cluster. The combination of the simplicity of linear models with the power of the statistical methods yields a compelling route for the deeper understanding of complex dynamics and we expect our approach to be widely applicable.

\section*{Methods}

\noindent{\bf Linear dynamics and the likelihood function:} We approximate a given time series using first-order linear dynamical systems in short windows and use a likelihood-ratio test to estimate whether new observations can be modeled by the linear coefficients. Given a $d$-dimensional discrete time series $\vec{x} \in \mathbb{R}^d$, we define the first order vector autoregressive process,

\begin{equation} \label{SEq:VAR}
\vec{x}_{t+1}=\vec{c}+\mathbf{A}\vec{x}_{t}+\vec{\eta}_{t+1},
\end{equation}
where $\vec{c} \in \mathbb{R}^d$ is an intercept vector, $\mathbf{A}$ is a $d\times d$ discrete time coupling matrix and $\vec{\eta}$ is a noise term with covariance $\Sigma$, which we assume to be Gaussian and white. We estimate the linear parameters $\theta = (\vec{c},\mathbf{A},\Sigma)$ through least squares regression. The continuous time linear couplings, $\phi$, can be obtained by taking

\begin{equation}\label{SEq:RealTimeCoef}
\phi=\frac{\mathbf{A}-\mathds{1}_d}{\Delta t},
\end{equation}
where $\mathds{1}_d$ is a $d$-dimensional identity matrix and $\Delta t$ is the inverse of the sampling rate.

 \noindent Using windowed data $X_{k+1} = \vec{x}_t, t\in [t_0,t_0+w_{k+1}]$ we construct the log-likelihood ratio between models with parameters $\theta_k$ and $\theta_{k+1}$ as
\begin{equation} \label{SEq:LogLikRatio}
\Lambda_{k,k+1} = l(\theta_{k+1}|X_{k+1}) - l(\theta_k|X_{k+1}).
\end{equation}
where the pseudo log-likelihood function of model parameters $\theta_a=\left(\vec{c}_a,\mathbf{A}_a,\Sigma_a\right)$ from $X_b$ for a Gaussian process is given by
\begin{equation}\label{SEq:LogLik}
l(\theta_a|X_b)= -\frac{1}{2}\sum_{t=t_0+1}^{w_b}\left\{\text{log}\left[(2\pi)^d|\Sigma_a|\right]+\vec{\eta}_t^\top\Sigma_a^{-1}\vec{\eta}_t\right\},
\end{equation}
where $\vec{\eta}_t$ is the error of modeling $X_b$ with $\theta_a$. 

\medskip 

\noindent{\bf Adaptive locally-linear segmentation algorithm:} We first define a set of candidate windows in which to examine whether there are dynamical breaks. This is done iteratively: we set a minimum window size $w_{min}$ and then increment by $\sim10\%$ which ensures that larger windows contain a proportionally larger number of observations. The candidate windows range between $w_{min}$ and some $w_{max}$ which corresponds to the value at which the step size is larger or equal to $w_{min}$. The specific value of $w_{min}$ depends on the dataset and the dimensionality $d$ and we chose $w_{min}$ to be the smallest interval in which the data can be reliably fit. However, simply setting $w_{min}=d$ does not incorporate the possibility of multicollinearity, when two or more components are not linearly independent, which produces an ill-conditioned linear regression. This linear dependence results in a moment matrix $\mathbf{X}^\top \mathbf{X}$ that is not full rank or nearly singular, and therefore small perturbations result in large fluctuations in the estimated linear parameters. In addition, computing the log-likelihood function \eqref{SEq:LogLik} requires inverting the covariance matrix of the error $\Sigma$. Thus, we require a minimum window size for which both $\mathbf{X}^\top \mathbf{X}$ and $\Sigma$ are well-conditioned. We compute the condition number of these matrices as a function of window size and choose $w_{min}$ as the smallest window for which the condition numbers are reasonably small. The results for each analyzed dataset are shown in Fig.\,(S8). 

\medskip

\noindent Given a set of candidate windows we iterate over pairs of consecutive windows of size $w_k$ and $w_{k+1}$, estimate the respective model parameters $\theta_k$ and $\theta_{k+1}$, and locate a dynamical break if $\theta_{k+1}$ performs significantly better than $\theta_k$ in fitting the data from the window of size $w_{k+1}$.  We assess significance through a likelihood ratio test and obtain $\Lambda_{k, k+1}$ from \eqref{SEq:LogLikRatio}.  We note that our models are non-nested
for which the likelihood ratio would be asymptotically $\chi^2$ distributed.  Instead, we take $\theta_{k}$ as a null model for the observations in the window of size $w_{k+1}$ and use a Monte Carlo approach to generate $N=5000$ surrogate trials of size $w_{k+1}$ from $\theta_k$  in order to compute $P_{\rm null}(\Lambda)$, the distribution of the log-likelihood ratio under the null hypothesis of having no model change.  We identify a dynamical break if $\Lambda_{k,k+1}>\Lambda_{\rm thresh}$ where $\Lambda_{\rm thresh}$ is defined by the larger solution of $P_{\rm null}(\Lambda_{\rm thresh})= 0.05$. A graphical representation of the technique is shown in Fig.\,(\ref{fig:Panel1}) and the algorithm is detailed in the Supplementary Information (SI). Finally, if the algorithm iterates to the maximum window size $w_\text{max}$ we automatically assign a break which we then asses through the following procedure: we start with $w_k=w_\text{min}$ and compare the models found in the intervals $[w_\text{max}-w_k,w_\text{max}]$ and $[w_\text{max}-w_k,w_\text{max}+(w_{k+1}-w_k)]$ as we increase $k$ until we span the entire set of candidate windows. If none of these tests suggest a break then we simply remove it. 

\medskip

\noindent We choose the significance threshold empirically and this choice reflects a tension between model complexity and accuracy; varying $P_{\rm null}(\Lambda_{\rm thresh})$ principally changes the number of breaks. While we have found  $P_{{\rm null}}(\Lambda_{{\rm thresh}})=0.05$ to be reasonable across multiple datasets we provide additional intuition through a toy segmentation problem illustrated in SI. The results reported in this manuscript do not depend sensitively on the significance threshold.

\medskip

\noindent{\bf Regularization for high-dimensional data:} Regularization can be incorporated straightforwardly into our method: at each iteration step we project the windows of size $w_k$ and $w_{k+1}$ to a space of orthogonal vectors defined by the first $D$ eigenvectors of the covariance matrix of the window of size $w_{k+1}$, estimate whether a break exists in this lower dimensional space, and then project back the inferred model parameters to the original space through a simple linear transformation. The number of eigenvectors is chosen to keep the condition number of the covariance matrix of the data and the error below a certain threshold $\kappa_\text{thresh}$ in order to ensure a well-conditioned model fit. We chose to use the condition number instead of the fraction of explained variance as a threshold, such that we can capture as much of the variance while being able to have a well-conditioned model fit. This results in projections that can capture more of the variance than that imposed by a variance threshold. We set the minimum window size at $10$ frames, such that $w_{k+1}$ is at least 10\% larger than $w_k$ (as in Algorithm 1 of the Supplementary Information). We demonstrate this regularization procedure on a dataset consisting of calcium imaging of hundreds of neurons in the mouse visual cortex, Figs.\,(6B,\,S7). Other approaches such as lasso or ridge regression may also be incorporated, but at the cost of additional regularization parameters  \cite{Hoerl1970,Santosa1986}.  

\medskip

\noindent{\bf Likelihood-based hierarchical clustering:} The space of linear dynamical systems has a family of equivalent representations given by the transformation $P \in \text{GL}(n)$ of the group of non-singular $n\times n$ matrices and thus the Euclidean metric is not an appropriate dissimilarity measure. While previous solutions have been presented for measuring LDS distances \cite{Afsari2013,Ravichandran2009}, the adaptation of these methods to our framework would be intricate and unnatural and we instead define a likelihood dissimilarity measure, which is consistent with the adaptive segmentation method. In essence, two models are distant if the model found by combining the two corresponding windows is unlikely to fit either window. On the other hand, when the models are similar, then the model found by combining the two windows is very likely to fit both windows. Specifically, let $\Lambda_{c,a} = l(\theta_a|X_a) - l(\theta_c|X_a)$ and $\Lambda_{c,b} = l(\theta_b|X_b) - l(\theta_c|X_b)$. We define the dissimilarity between models $\theta_a$ and $\theta_b$ as,
\begin{equation}\label{SEq:LikDistance}
d_{a,b} = \Lambda_{c,a} + \Lambda_{c,b},
\end{equation}
where $X_c = X_a \cup X_b$, and $\theta_c$ is the result of fitting $X_c$ to \eqref{SEq:VAR}. This measure is positive semi-definite since $\Lambda_{c,a}  \geq 0$ and
 $\Lambda_{c,b}  \geq 0$ ($\theta_a$ is the maximum likelihood estimate in $X_a$; $l(\theta_a|X_a) - l(\theta_c|X_a) \geq 0$) and also symmetric; since we do not fit across windows, a first order linear fit in $X_a \cup X_b$ yields the same linear couplings as in $X_b \cup X_a$.  After computing the dissimilarity between all linear models, we use Ward's criterion \cite{Ward1963} to perform hierarchical clustering by minimizing the within-cluster variance. 

\medskip

\noindent{\bf  Lorenz data:} We simulated the Lorenz system using the scipy.odeint package \cite{Scipy} with parameter choices $\sigma=10$, $\beta=8/3$ and $\rho=28$ in the chaotic regime and $\rho=20$ for spirals. We used step size $\Delta t=0.02 \, {\rm s}$. In the chaotic regime we integrated for a total of 1000\,s, waiting 200\,s for the trajectories to fall onto the attractor. For the stable spirals in the late-time transient chaos regime, we chose initial conditions $(x_0,y_0,z_0)=(x,0,20)$, where $x$ varied from $-12$ to $-8$ and $8$ to $12$ in steps of $0.2$, yielding a total of $42$ initial conditions. The trajectories were drawn to one of the stable fixed points $C_{\pm}=(x^*,y^*,z^*)=(\pm\sqrt{\beta(\rho-1)},\pm\sqrt{\beta(\rho-1)},\rho-1)$, for which linear stability analysis yields a stable oscillation with $\lambda_r\approx -0.4$ and $\lambda_i/(2\pi)\approx1.4$ and a relaxation with $\lambda_r\approx -12.9$. We waited for 10\,s before sampling the spiraling trajectory on additional 10\,s. To reduce multi-collinearity we added small amplitude Gaussian white noise with a diagonal covariance matrix with variances $\sigma_{ii} = 0.001, \hspace{0.2cm} i \in  \{x,y,z\}$ to the simulated time series. The minimum window size  $w_{min}=10$ frames was chosen using Fig.\,(S8).

\medskip 

\noindent{\bf {\em C. elegans} posture data:}  We analyzed previously published data consisting of foraging behavioral conditions \cite{Stephens2011,Stephens2008} in which N2-strain \textit{C.~elegans}   were imaged at $f=32\,{\rm Hz}$ with a video tracking microscope.
Coiled shapes were resolved and the time series downsampled to $f=16\,{\rm Hz}$ \cite{Broekmans2016}. Worms were grown at $20{^\circ C}$ under standard conditions \cite{Sulston1974}. Before imaging, worms were removed from bacteria-strewn agar plates using a platinum worm pick, and rinsed from \textit{E.~coli} by letting them swim for $1\, {\rm min}$ in NGM buffer. They were then transferred to an assay plate ($9\, {\rm cm}$ Petri dish) that contained a copper ring ($5.1\, \rm{cm}$ inner diameter) pressed into the agar surface, preventing the worm from reaching the side of the plate. Recording started approximately $5\, \rm{min}$ after the transfer, and lasted for $2100\, \rm{s}$. In total, data from $N=12$ worms was recorded. Using Fig.\,(S8), we selected a minimum window size of $w_{min}=10$ frames.  Likelihood hierarchical clustering yielded a dendrogram for which a cut at the 4-branch level resulted in clusters with approximately 6500 (fast forward), 14400 (slow forward), 3500 (turns) and 4200 (reversals) models. In Fig.\,(\ref{fig:Panel4}B), reversal events were identified when the phase velocity changes sign. Only segments for which there is a 2\,s window of positive and negative phase velocity before and after the change of sign are considered.

\medskip

\noindent{\bf {\em C. elegans} neural data:}  We analyzed whole-brain experiments from the Zimmer group in which transgenic {\em C.~elegans} expressing a nuclear localized $\text{Ca}^{2+}$ indicator were imaged in a microfluidic device where a reduction in $\text{O}_2$ concentration was observed to induce a ``sleep''-like, quiescent state in \emph{npr}-1 lethargus animals \cite{Nichols2017}. A range of 99-126 neurons was imaged for $N=11$  worms and each neural trace was normalized by subtracting the background and dividing by the mean signal. A linear component was also subtracted to correct for bleaching. We used principal components analysis to reduce each ensemble recording to an 8-dimensional time series capturing $\sim90\%$ of the variance. Each of the experimental trials (one per worm) consisted of three 6 minute periods with alternating $\text{O}_2$ concentrations: starting with $10\%$, increasing to $21\%$ and returning to $10\%$.  We selected a minimum window size $w_{min}=18$ frames using Fig.\,(S8). Likelihood hierarchical clustering yielded a dendrogram for which a cut at the 3-branch level resulted in one cluster with  24 models, another with 74 models and a third outlier cluster containing just 1 model. Removing the outlier resulted in a dendrogram with a more even model distribution: one cluster with 24, another with 16 and a third with 55 models.   We used this clustering to compare state labels with ``active'' and ``quiescent'' global brain states identified previously \cite{Nichols2017}, Fig.\,(S4). Data from unperturbed worms exhibiting more complex natural behaviors \cite{Kato2015} was analyzed similarly.

\medskip

\noindent{\bf {Monkey electrocorticography data:}} We analyzed a publicly available dataset (\url{http://neurotycho.org/}), that was previously described \cite{Solovey2015,Tajima2015,Yanagawa2013}. The details of the experimental procedure can be found in \cite{Nagasaka2011}. The raw 128 electrode signals were preprocessed in the following way. First, the original signal was downsampled from 1\,KHz to 500\,Hz. Then, two channels were removed due to significant line noise contamination. The remaining 126 electrodes were filtered to remove the line noise at 50\,Hz and subsequent harmonics. Multi-taper filtering was performed using the Chronux toolbox \cite{Mitra2008}, available at \url{http://chronux.org/}, with a bandwidth of 5\,Hz (9 tapers) in a moving window of 2\,s with 0.5\,s overlap. The overlap regions were smoothed using a sigmoid function with smoothing parameter $\tau=10$. Finally, the electrode signals were projected into 40 principal components that capture $\sim 99\%$ of the variance. We selected a minimum window size $w_{min}=83$ frames using Fig.\,(S8).

\medskip

\noindent{\bf \em{Mus musculus} neural data:} We analyzed a publicly available dataset (\url{http://observatory.brain-map.org/visualcoding/search/cell_list?experiment_container_id=511854338&sort_field=p_sg&sort_dir=asc}) from the Allen Institute \cite{AllenBrainObservatory}. The analyzed data constituted a total of 240 neurons from the anterolateral visual cortex of \emph{Mus musculus}, at a depth of $275\,\mu\rm{m}$. Neural activity was sampled at $\sim 30\,\rm{Hz}$ for $\sim 60\,\rm{mins}$ with a GCaMP6f calcium indicator, during exposure to a natural movie. The background subtracted bleach corrected signals were accessed using the Allen Software Development kit \cite{AllenSDK}. The local linear analysis was performed with regularization using a condition number threshold of $\kappa_\text{thresh}=10^5$.

\medskip

\noindent{\bf Software:} Code for the adaptive locally-linear segmentation and likelihood-based hierarchical clustering was written in Python \cite{Python} and is publicly available (\url{https://github.com/AntonioCCosta/local-linear- segmentation.git}).

\section*{Acknowledgements}

We thank SURFsara (www.surfsara.nl) for computing resources through the Lisa system. This work was supported by the research program of the Foundation for Fundamental Research on Matter (FOM), which is part of the Netherlands Organization for Scientific Research (NWO), and also by funding from The Okinawa Institute of Science and Technology Graduate University. GJS also acknowledges useful discussions at the Aspen Center for Physics, which is supported by National Science Foundation grant PHY-1607611.

% \noindent 

\clearpage

\bibliographystyle{apsrev4-1_PRX_style}
\bibliography{Bibliography}

\clearpage

\onecolumngrid
\setcounter{figure}{0}
% \setcounter{page}{1}

% \widetext
% \begin{center}
% \textbf{\Large Supplementary Material}
% \end{center}
\section*{Supplementary Material}

%\makeatletter
\renewcommand{\theequation}{S\arabic{equation}}
\renewcommand{\thefigure}{S\arabic{figure}}

\section*{ Adaptive locally-linear segmentation algorithm} 

\begin{figure}
\begin{center}
\includegraphics{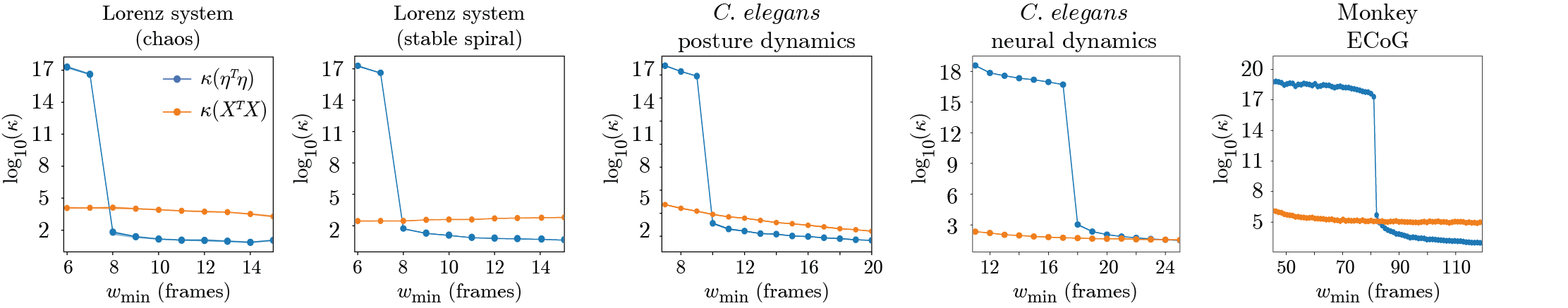}
\caption{{\bf Condition number of the moment matrix $\mathbf{X^\top X}$ and the error covariance matrix $\mathbf{\eta^\top\eta}$ as a function of window size.} We select the minimum window size as the smallest number of frames for which the model fit and log-likelihood estimation are well-conditioned. The condition number of $X^\top X$ and $\eta^\top\eta$ is calculated for different window sizes, and the median is estimated across samples taken randomly at different times in the time series. There is a drastic decrease in the condition number beyond a minimum window whose size depends on the data.  Beyond this window, the model fit and log-likelihood estimation are well-conditioned. 
}
\label{Fig:S_condnumber}
\end{center}
\end{figure}

We first define a set of candidate windows in which to examine whether there are dynamical breaks. This is done iteratively: we set a minimum window size $w_{min}$ and then increment by $\sim10\%$ which ensures that larger windows contain a proportionally larger number of observations. The candidate windows range between $w_{min}$ and some $w_{max}$ which corresponds to the value at which the step size is larger or equal to $w_{min}$. The specific value of $w_{min}$ depends on the dataset and the dimensionality $d$ and we chose $w_{min}$ to be the smallest interval in which the data can be reliably fit. However, simply setting $w_{min}=d$ does not incorporate the possibility of multicollinearity, when two or more components are not linearly independent, which produces an ill-conditioned linear regression. This linear dependence results in a moment matrix $\mathbf{X}^\top \mathbf{X}$ that is not full rank or nearly singular, and therefore small perturbations result in large fluctuations in the estimated linear parameters. In addition, computing the log-likelihood function Eq.~(4) requires inverting the covariance matrix of the error $\Sigma$. Thus, we require a minimum window size for which both $\mathbf{X}^\top \mathbf{X}$ and $\Sigma$ are well-conditioned. We compute the condition number of these matrices as a function of window size and choose $w_{min}$ as the smallest window for which the condition numbers are reasonably small. The results for each analyzed dataset are shown in Fig.\,(\ref{Fig:S_condnumber}).

\begin{algorithm}[htp]
\caption{ Iterative construction of window sizes}
\label{alg:algorithm1}
\begin{algorithmic}
\STATE{$w = w_{min}$} 
\STATE{$s = 0$}
\WHILE{$s < w_{min}$}
	\STATE{save $w$}
    \STATE{$s= \text{int}(w/10)$}
    \IF{$s<w_{min}$ }
        \STATE{$w = w + s$}
    \ELSE{}
    \STATE{break}
    \ENDIF
\ENDWHILE
\end{algorithmic}
\end{algorithm}

Given a set of candidate windows we iterate over pairs of consecutive windows of size $w_k$ and $w_{k+1}$, estimate the respective model parameters $\theta_k$ and $\theta_{k+1}$, and locate a dynamical break if $\theta_{k+1}$ performs significantly better than $\theta_k$ in fitting the data from the window of size $w_{k+1}$.  We assess significance through a likelihood ratio test and obtain $\Lambda_{k, k+1}$ from Eq.~(3).  We note that our models are non-nested
for which the likelihood ratio would be asymptotically $\chi^2$ distributed.  Instead, we take $\theta_{k}$ as a null model for the observations in the window of size $w_{k+1}$ and use a Monte Carlo approach to generate $N=5000$ surrogate trials of size $w_{k+1}$ from $\theta_k$  in order to compute $P_{\rm null}(\Lambda)$, the distribution of the log-likelihood ratio under the null hypothesis of having no model change.  We identify a dynamical break if $\Lambda_{k,k+1}>\Lambda_{\rm thresh}$ where $\Lambda_{\rm thresh}$ is defined by the larger solution of $P_{\rm null}(\Lambda_{\rm thresh})= 0.05$. A graphical representation of the technique is shown in Fig.\,(1) and the algorithm is detailed below. Finally, if the algorithm iterates to the maximum window size $w_{max}$ we automatically assign a break which we then asses through the following procedure: we start with $w_k=w_{min}$ and compare the models found in the intervals $[w_{max}-w_k,w_{max}]$ and $[w_{max}-w_k,w_{max}+(w_{k+1}-w_k)]$ as we increase $k$ until we span the entire set of candidate windows. If none of these tests suggest a break then we simply remove it. 

\begin{figure}
\begin{center}
\includegraphics{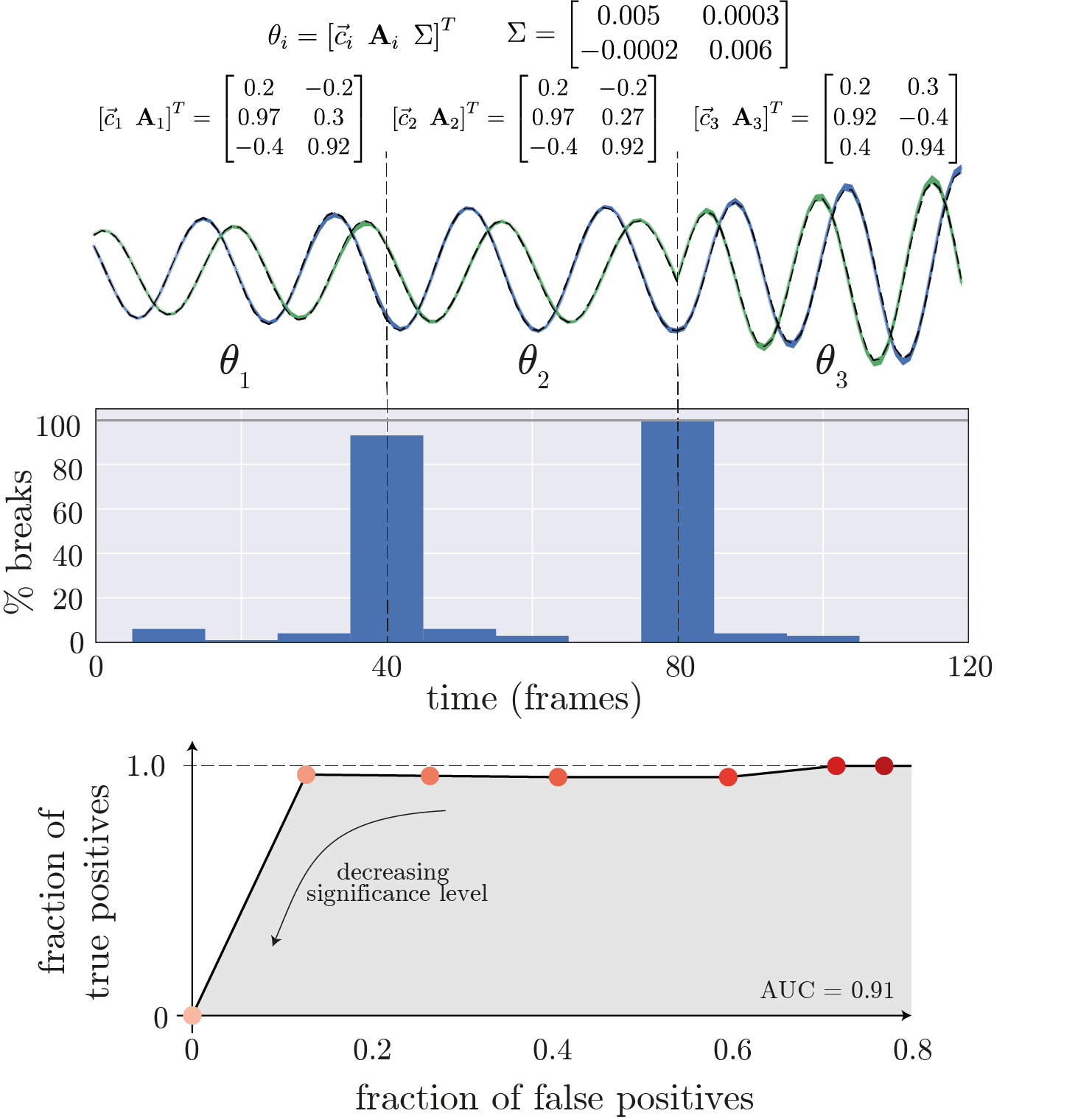}
\caption{{\bf Accuracy of the adaptive segmentation technique on a 3-state toy example.} (top) We generate a set of 100 toy time series, for which there are three dynamical regimes, $\{\theta_1,\theta_2,\theta_3\}$. We plot a sample time series (dashed line) as well as the mean of the simulations made from the models and windows found using the adaptive segmentation technique (blue and green lines represent $x_1$ and $x_2$, respectively; shade represents bootstrapped 95\% confidence intervals). (middle) We plot the distribution of dynamical breaks, across simulations, for the smallest significance level $P_{\rm null}(\Lambda_{\rm thresh})=0.01$, for which both breaks are found with high accuracy. (bottom) Fraction of true positives (breaks found by the algorithm that correspond to dynamical changes) and fraction of false positives (breaks that the algorithm found that do not correspond to dynamical changes) as a function of the significance level: $\{60\%,40\%,20\%,10\%,5\%,2.5\%,1\%\}$ (dark red to light red represents decreasing significance levels). At high significance levels, the segmentation algorithm is very sensitive and thus the null hypothesis is rejected easily resulting in a large amount of false positives. As we decrease the significance level, we start rejecting the null hypothesis less, while still capturing the true dynamical changes. For significance levels below $5\%$, the fraction of false positives drops below $50\%$ while the fraction of true positives remains close to $100\%$. Indeed, the area under the curve (AUC) is nearly 1 and this is indicative of the quality of the segmentation.
}
\label{Fig:S_toy}
\end{center}
\end{figure}

We choose the significance threshold empirically and this choice reflects a tension between model complexity and accuracy; varying $P_{\rm null}(\Lambda_{\rm thresh})$ principally changes the number of breaks. While we have found  $P_{{\rm null}}(\Lambda_{{\rm thresh}})=0.05$ to be reasonable across multiple datasets we provide additional intuition through a toy segmentation problem illustrated in Fig.\,(\ref{Fig:S_toy}). We simulate $N=100$ two-dimensional systems $\vec{x}_s$ for which we change the model parameters twice: first we apply a small change to the coupling between $x_1$ and $x_2$, $A_{12}\rightarrow A_{12}+0.03$, while in the next change we symmetrize the couplings between $x_1$ and $x_2$ thus reverting the direction of the oscillation, Fig.\,(\ref{Fig:S_toy}-top). Both change points are accurately determined even for a significance level of $1\%$, Fig.\,(\ref{Fig:S_toy}-middle); the dynamical changes are found $\sim 96\%$ of the time, even though the change between the first two models is quantitatively small.  In Fig.\,(\ref{Fig:S_toy}-bottom) we show the number of true positives (breaks found by the algorithm that are true dynamical changes) and false positives (breaks found by the algorithm that are not true dynamical changes) as a function of the significance level. The fraction of true positives is essentially preserved (even if we stretch to a $1\%$ significance level), indicating that missing true dynamical changes is rare. The results reported in this manuscript do not depend sensitively on the significance threshold.

\begin{algorithm}[htp]
\caption{Description of the adaptive locally-linear segmentation of a $d$-dimensional time series $\vec{x}$, of length $T$ given a set of $N_w$ candidate windows.}
\label{alg:algorithm2}
\begin{algorithmic}
\STATE{$t=0$} 
\WHILE{$t < T$}
	\STATE{$k=0$ }
    \WHILE{k $<$ $N_w$}
    \STATE{$X_k=\vec{x}_t, t\in[t,t+w_k]$ }
    \STATE{$X_{k+1}=\vec{x}_t, t\in[t,t+w_{k+1}]$ }
    \STATE{Fit $\theta_k$ and $\theta_{k+1}$ to $X_k$ and $X_{k+1}$, respectively}
    \STATE{Compute $\Lambda_{k,k+1}$ over $X_{k+1}$, from $\theta_k$ and $\theta_{k+1}$} 
	\STATE{Generate $N_s$ time series $X^s_{k+1}$ of size $w_{k+1}$, using $\theta_k$}
	\STATE{Compute $\Lambda^s_{k,k+1}$ with the newly obtained $\theta^s_k$ and $\theta^s_{k+1}$, for each $N_s$ time series, obtaining a distribution $P_{\rm null}(\Lambda)$}
	\STATE{Estimate $\Lambda_\text{thresh}$ as the largest solution of $P_{\rm null}(\Lambda_\text{thresh})=0.05$ }
    \IF{$\Lambda_{k,k+1} \leq \Lambda_\text{tresh}$}
    \STATE{There is no dynamical change}
    \STATE{$k=k+1$}
    \ELSE{}
    \STATE{save the window $[t,t+w_k]$}
    \STATE{\bf break}
    \ENDIF
\ENDWHILE
\STATE{$t=t+w_k$}
\ENDWHILE
\end{algorithmic}
\end{algorithm}

\clearpage

%%% Each figure should be on its own page

\begin{figure}[htp]
\begin{center}
\includegraphics{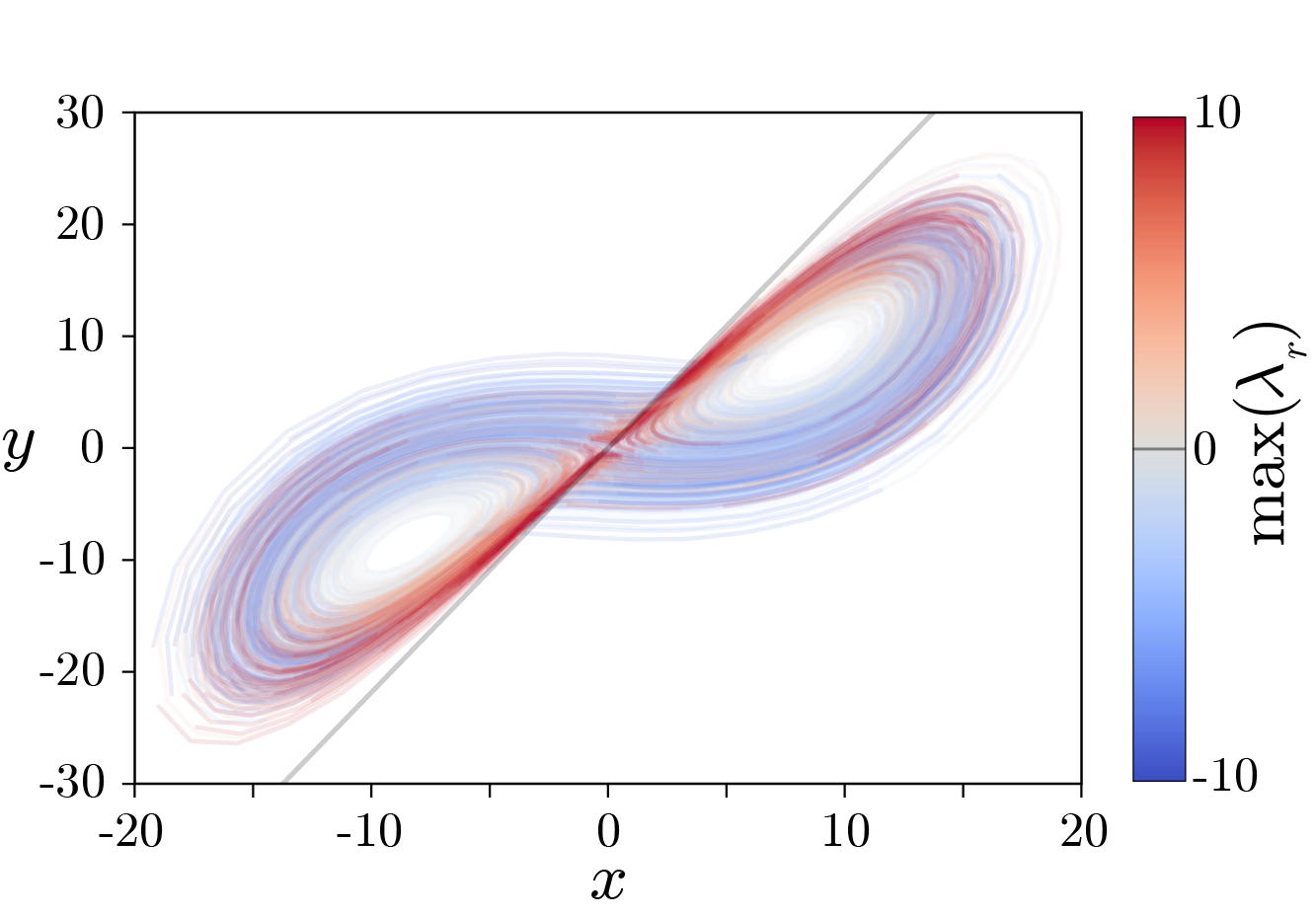}
\caption{{\bf Unstable linear models in the chaotic regime of the Lorenz system lie along the unstable manifold of the origin.} We plot the $xy$ projection of the Lorenz system in the chaotic regime, color coded by the magnitude of the maximum real eigenvalue $\lambda_r$ of the linear model found in each window resulting from the adaptive segmentation. Most unstable models are found close to the origin, along its $1d$ unstable manifold (gray line).
}
\label{Fig:S_LorenzUnstable}
\end{center}
\end{figure}

\begin{figure}
\begin{center}
\includegraphics{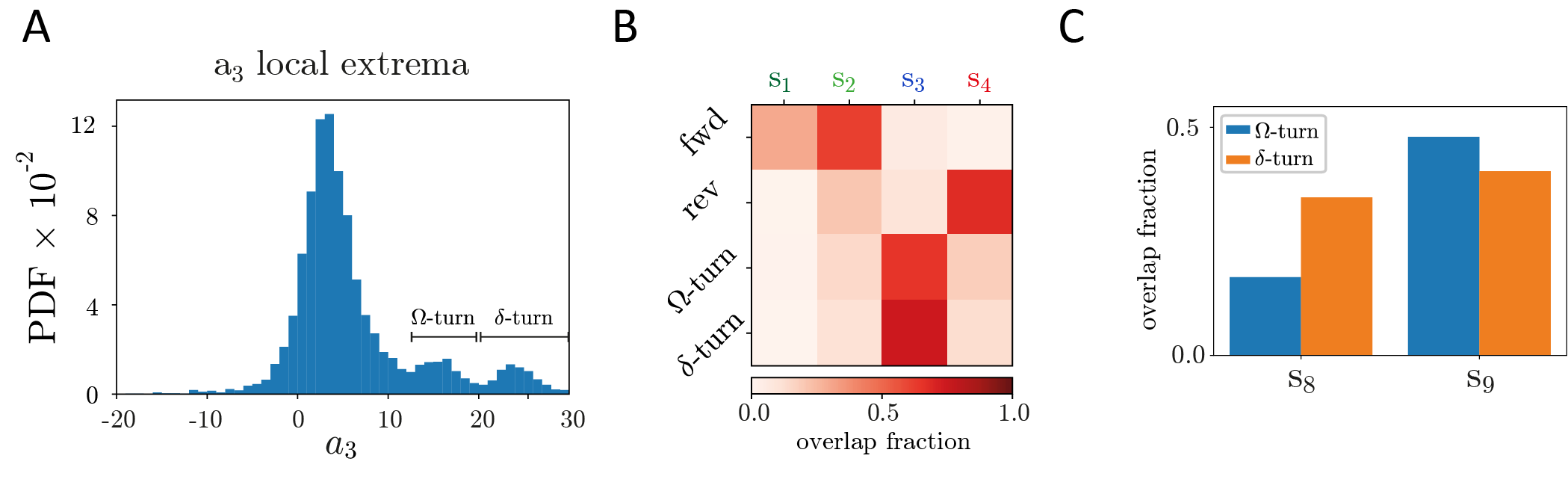}
\caption{{\bf Overlap between the clusters found in \emph{C. elegans} postural dynamics and behavioral motifs defined phenomenologically.} 
(A) We identified $\Omega$ and $\delta$ turns using similar criteria as \cite{Broekmans2016}. First, we found local extrema using scipy.signal.find\_peaks package \cite{Scipy} in Python \cite{Python}, with a prominence of 0.5. Only peaks without local extrema in a 3\,s window around are taken into account. For each of the windows identified with our locally-linear segmentation, we first check whether there is a well defined turn, according to the previous criteria. Then, if the amplitude of the peak is between 12 and 20, we classify it as an $\Omega$-turn; if it is larger than 20, we classify it as a $\delta$-turn. Windows for which there is no well defined turn and the maximum turning amplitude is below 12 are classified as either forward or reversal based on the sign of the phase velocity $\omega$. All remaining windows receive no label.
(B) Fraction of windows classified phenomenologically that fall into each of our clusters at a 4-branch level. The sparsity of the overlap matrix indicates that our clusters agree with classical definitions of coarse-grained behaviors.
(C) At a 12-branch level, cluster $s_8$ exhibits more $\delta$ that $\Omega$ turns, while $s_9$ exhibits more $\Omega$ turns. Nonetheless, there is still some confusion between $\delta$ and $\Omega$ turns. Our local linear models span time scales that are shorter than a typical turn, thus subdividing it. Therefore it is not surprising that there is some confusion between $\delta$ and $\Omega$ turns at this level of clustering: the only distinction between these is the height of the $a_3$ peak, which will correspond to only a small fraction of linear models. In addition, our clustering takes into account the entire dynamical pattern and therefore the value of the peak alone plays a minor role in differentiating between clusters. 
}
\label{Fig:S_overlapPNAS}
\end{center}
\end{figure}

\begin{figure}
\begin{center}
\includegraphics{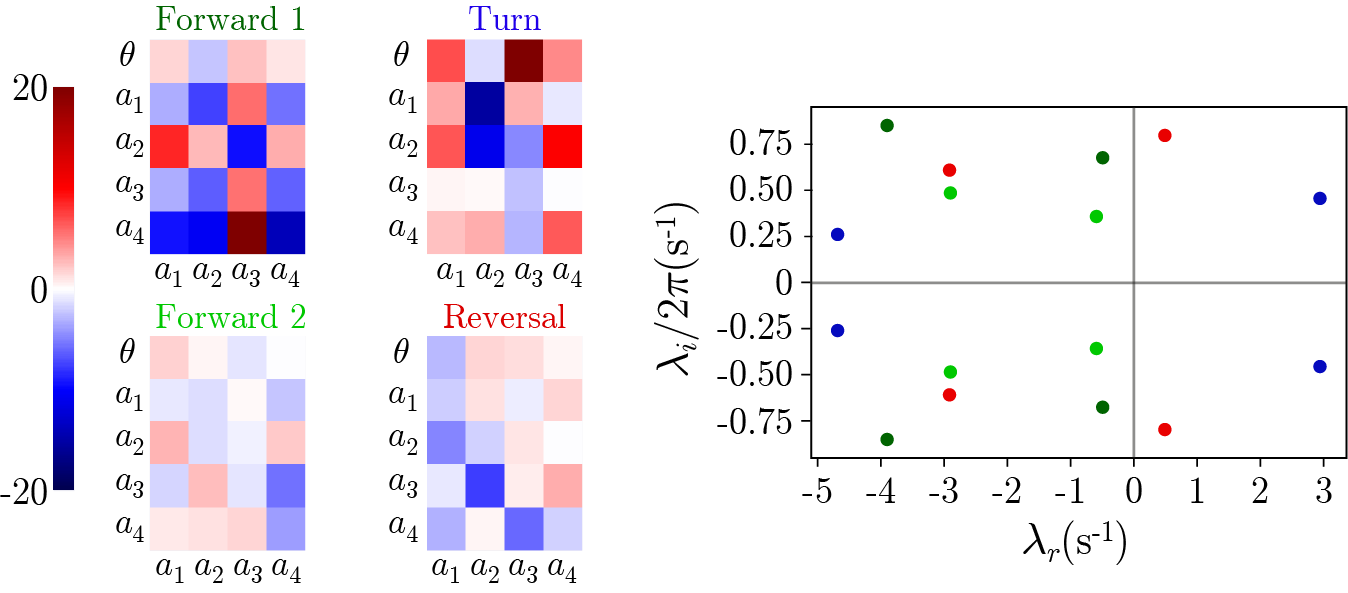}
\caption{{\bf Distinct behavioral classes in the dendrogram interpreted through model parameters.} 
The mean $\vec{\mu}$ and linear couplings $\phi$  displayed as a matrix $\left(\vec{\mu},\hspace{0.2cm}\phi\right)^\top$ (left) and the respective dynamical eigenvalues (right) are shown for a set of example models. From the first to the second forward state, the imaginary eigenvalues shrink, corresponding to a reduction of the oscillation frequencies.  The turn state exhibits a higher value of the mean of $a_3$ and, in this example at the beginning of a turn, we find an unstable oscillation. Finally, in the reversal state, the sign of the coupling between the first two modes is reversed and this signals a change in the sign of the phase velocity.
}
\label{Fig:S_ExampleModels}
\end{center}
\end{figure}

\begin{figure}[htp]
\begin{center}
\includegraphics{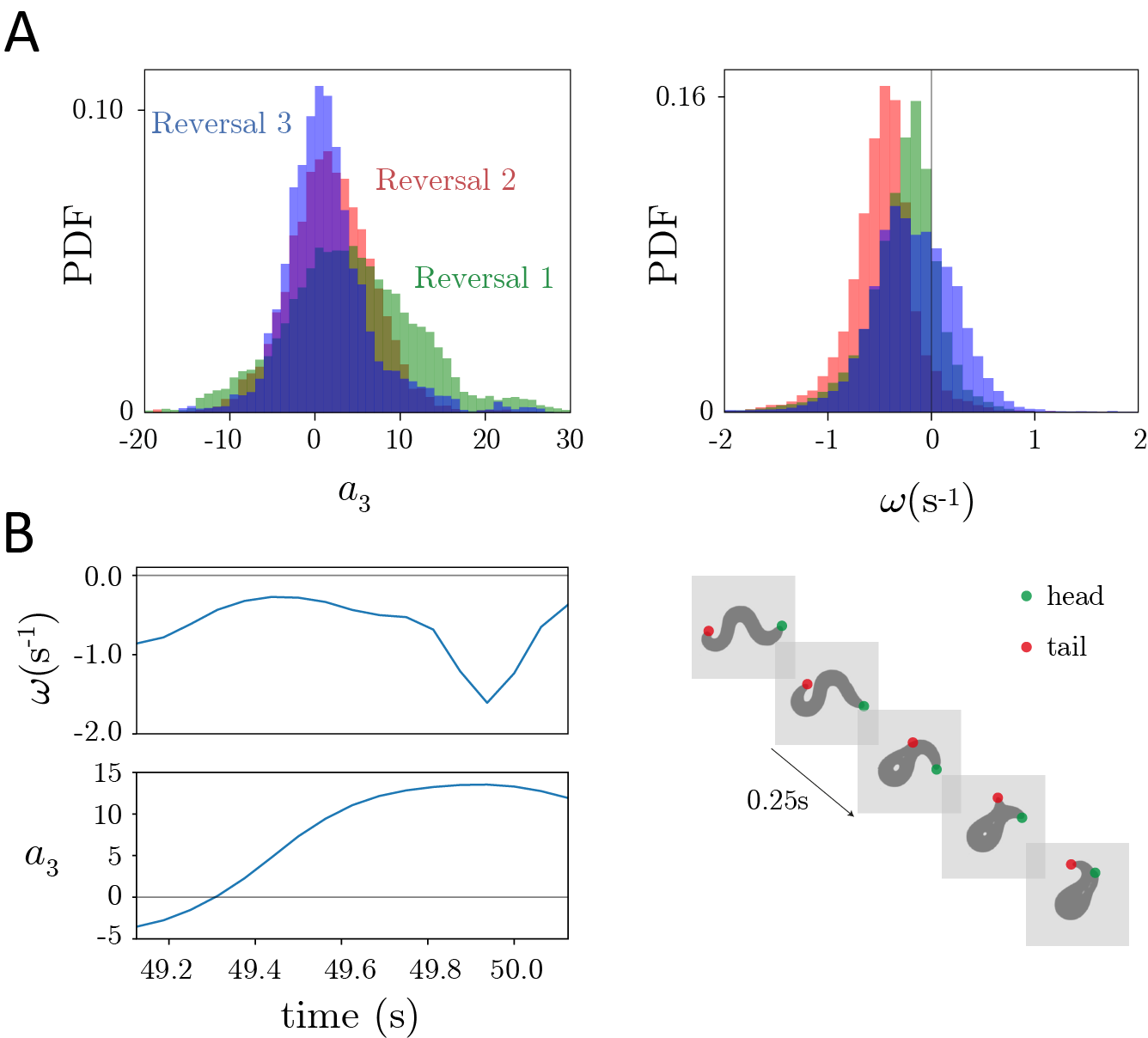}
\caption{{\bf At a 12-branch level of the dendrogram the reversal branch exhibits a reversal-turn behavioral motif.}
(A-left) Distribution of the turning amplitudes $a_3$. The first reversal (Reversal 1 - green) is actually a reversed turn, as noted by the high value of $a_3$. The other two reversals generally have smaller turning amplitudes. 
(A-right) Distribution of body wave phase velocities $\omega$. The second reversal branch (Reversal 2 - red) corresponds to faster reversal bouts, while the third reversal (Reversal 3 - blue) includes movements at the start of a reversal when $\omega$ is small.
(B-left) Example trajectory of a reversal-turn. A negative phase velocity $\omega$ is accompanied by a peak in $a_3$ for which the body is bent as in an $\Omega$-turn.
(B-right) Worm images from the example trajectory sampled each 0.25~s. The head and tail are identified with a green and red dot, respectively.
}
\label{Fig:S_reversals}
\end{center}
\end{figure}

\begin{figure}[htp]
\begin{center}
\includegraphics{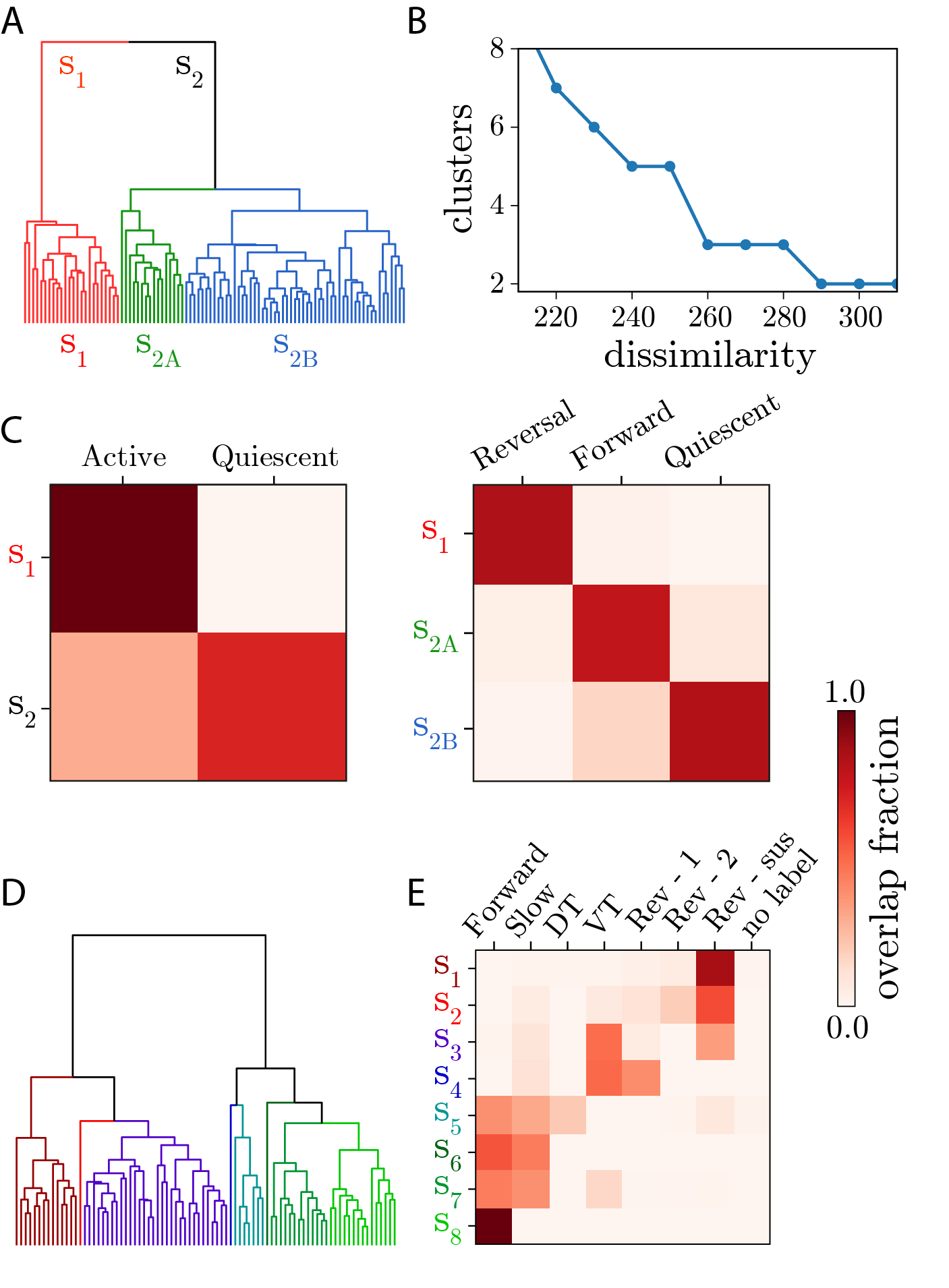}
\caption{{\bf  Model clusters exhibit approximate correspondence with labeled brain states.}
(A) Model clustering dendrogram obtained by segmenting the neural activity of the example worm shown in Fig.\,(5A).
(B) Number of clusters as a function of dissimilarity. The first major splits occur at the two branch and three branch level.
(C) Overlap between model clusters and labeled brain states \cite{Nichols2017}. At the two branch level, we find that most of the frames in branch $S_1$ were labeled as ``active'', while frames in branch $S_2$ were labeled as ``quiescent''.  At the three branch level we find a high degree overlap between $S_1$ and ``reversal'', $\text{S}_{2\text{A}}$ and ``forward'', and $\text{S}_{2\text{B}}$ and ``quiescent''.
(D)  Model clustering dendrogram obtained by segmenting the neural dynamics of an exemplar worm from previous global brain imaging experiments \cite{Kato2015}.
(E) Overlap between model clusters and labeled brain states from previous experiments \cite{Kato2015}. The sparseness in the matrix indicates a broad match between the states, specially for the forward and reverse states. (DT - Dorsal Turns, VT - Ventral Turns, Rev-Sus - Sustained Reversal)
}
\label{Fig:S_sleep}
\end{center}
\end{figure}

\begin{figure}
\begin{center}
\includegraphics{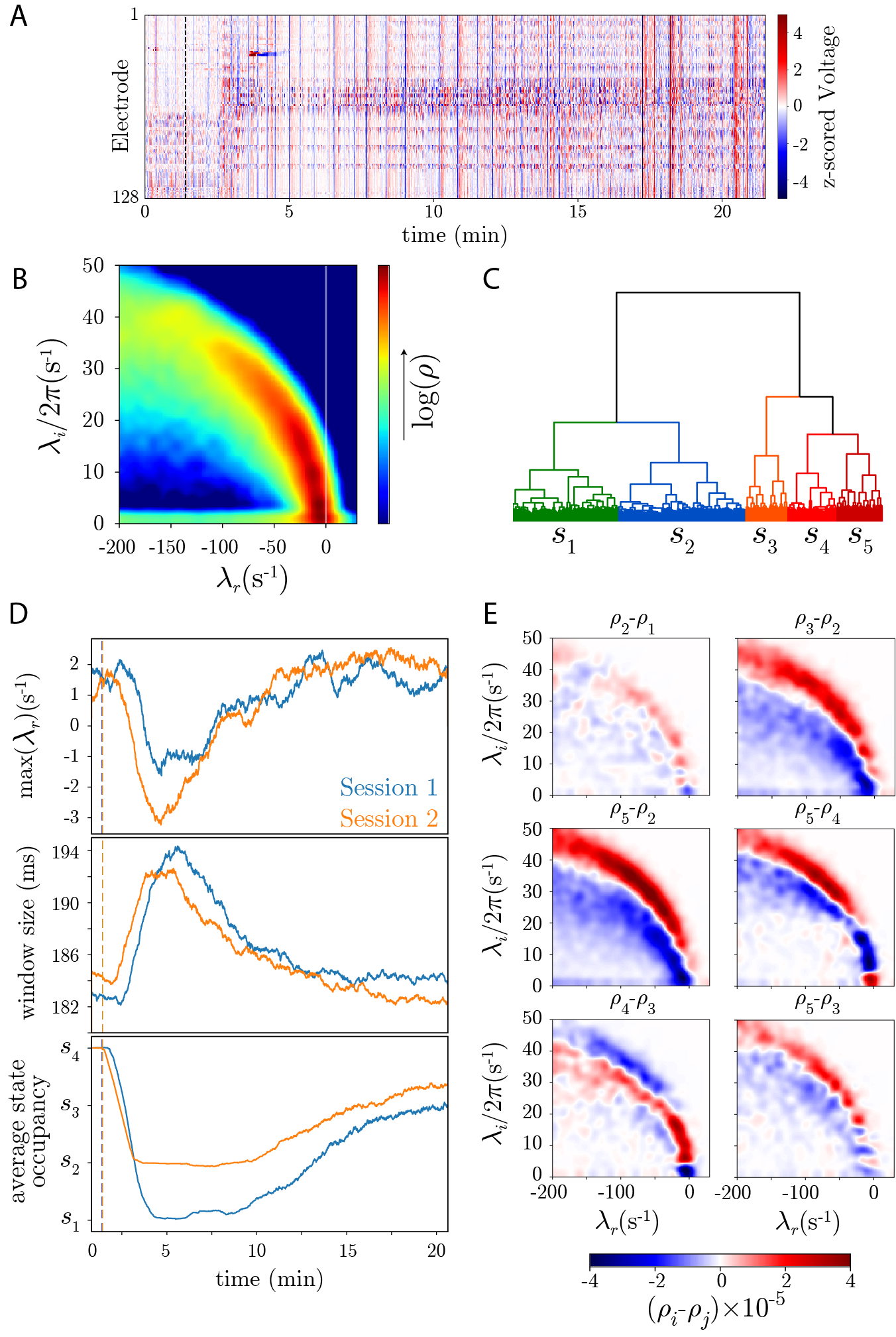}
\caption{{\bf Locally-linear analysis in the higher-dimensional  context of electrocorticography recordings from non-human primates.} 
(A) Example traces from an experiment in which the injection of propofol (dashed line) induces anesthesia in the monkey subject. There are two trials which occurred with the same subject but on different days. We project the time series into a 40-dimensional space through principal component analysis, capturing $\sim 99\%$ of the variance.
(B) Eigenvalue spectrum of the collection of linear models found through adaptive segmentation. The distribution of eigenvalues spans a wide range of frequencies and peaks near the instability boundary.
(C) Dendrogram of the likelihood clustering of the space of models.
(D) Propofol injection (dashed line) induces profound changes in brain dynamics. (top) The injection first results in dynamics that are increasingly unstable, then more deeply stable followed by a slow relaxation towards the instability boundary. (middle) These effects are also present in the window sizes, which increase after the injection of the anesthetic drug, reflecting a period in which the dynamics is less nonlinear. (bottom) Anesthesia also results in an abrupt change in the average state occupancy. The two different sessions differ in their average anesthetized state: while in session 1 the dynamics sits more in $s_1$, in session 2 we find a higher occupancy of $s_2$. The curves were smoothed using a $1\,\rm{s}$ running average.
(E) Difference between the eigenvalue distributions of different clusters. In general, the clusters exhibit frequency dependent changes in stability. In $s_1$, higher frequency states are more damped, while frequencies in the $\delta$ band are long-lived. In contrast, $s_2$ exhibits less damped frequencies (specially in the $\theta$ and $\alpha$ bands). Compared to any other cluster however, both $s_1$ and $s_2$ have their high frequency dynamics significantly more damped. This loss of power in the $\beta$ band has been associated with loss of consciousness \cite{Ishizawa2016,Chauvette2011} and is naturally captured by our technique.}
\label{Fig:ECoG}
\end{center}
\end{figure}

\begin{figure}
    \centering
    \includegraphics{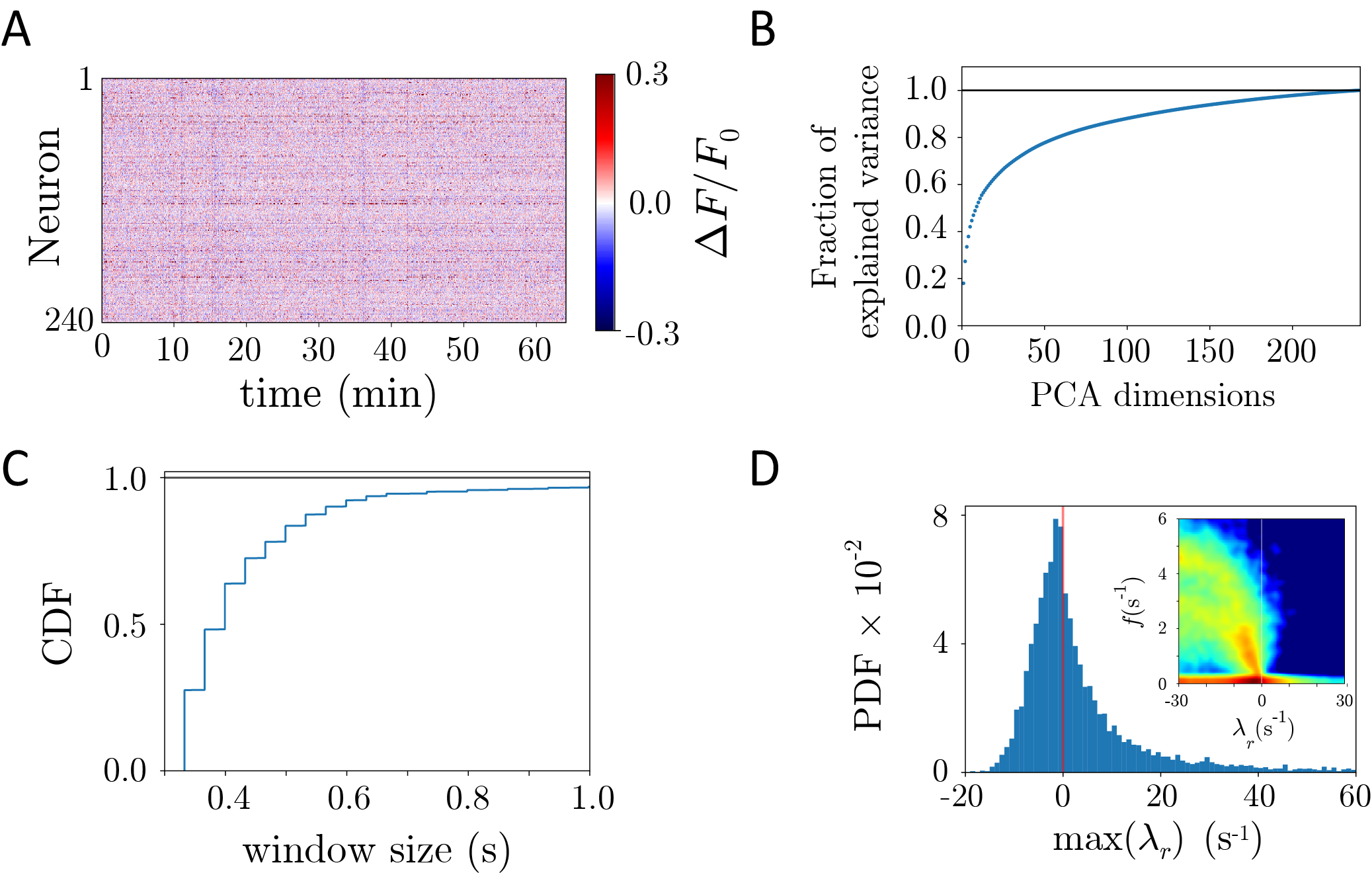}
    \caption{{\bf Locally-linear analysis with regularization applied to recordings of \emph{Mus musculus} visual cortex at single-cell resolution.} (A) Time series of 240 neurons from mouse visual cortex under a natural movie stimulus. (B) The neural population does not appear low dimensional: the spectrum of eigenvalues of the covariance matrix indicates that in order to capture most of the variance we would need almost as many principal components as the original number of cells. (C) Extending the original locally-linear analysis to include a regularization step, the inferred window sizes exhibit a wide distribution with heavy tails extending from $0.3\,\rm{s}$ to longer than $1\,\rm{s}$. Without regularization, the minimum window size is $\sim 500$ frames ($\sim 16.5\,\rm{s}$) in order to ensure a well-conditioned model fit. Here, we have used a condition number threshold of $\kappa_\text{thresh}=10^5$. Further details are given in Methods. (D) The neural dynamics exhibits a wide range of frequencies and dynamics that sit near the instability boundary. 
    }
    \label{Fig:AllenPCR}
\end{figure}

\

\begin{figure}
\begin{center}
\includegraphics{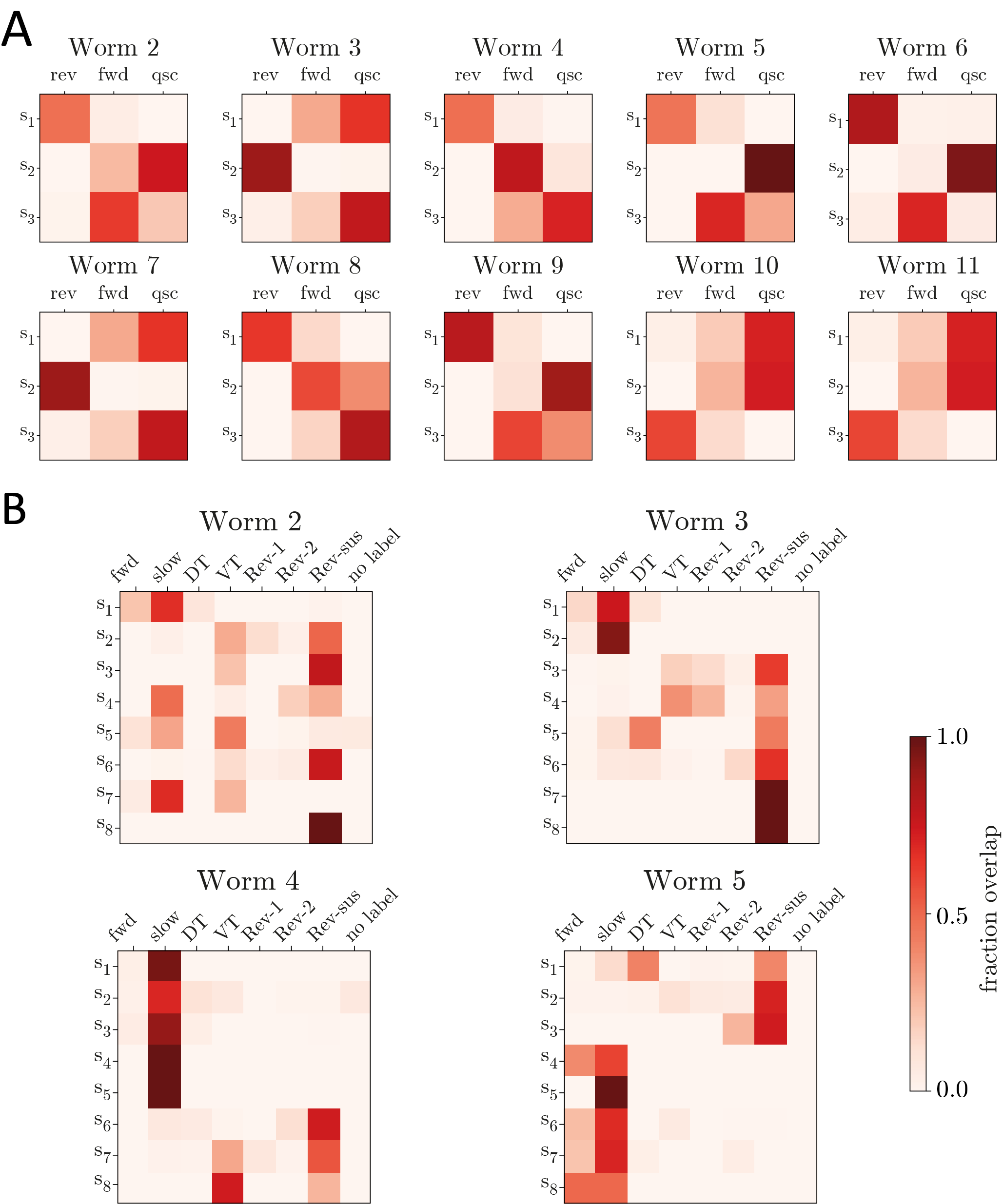}
\caption{{\bf Confusion matrices for all the worms not shown in Fig.\,(S5).} 
(A) - Induced quiescence experiments \cite{Nichols2017}.  As with the worm in Fig.\,(S5), we had to remove an outlier in worm 4. (rev - reversal, fwd - forward, qsc - quiescent)
(B) - No stimulus experiments \cite{Kato2015}. The sparsity of the confusion matrices indicates a large degree of overlap. (fwd - forward, slow - slow forward, DT - dorsal turn, VT - ventral turn, Rev-1 - reversal 1, Rev-2 - reversal 2, Res-sus - sustained reversal)
}
\label{Fig:S_overlap}
\end{center}
\end{figure}

\end{document}